\documentclass[11pt]{article}
\pdfoutput=1
\usepackage[T1]{fontenc}
\usepackage{lmodern}
\usepackage{slantsc}
\usepackage[protrusion=true,expansion=true]{microtype}
\usepackage{breakcites}
\usepackage{amsmath,amssymb,amsfonts,amsthm}
\usepackage{subcaption}
\usepackage{graphicx}
\usepackage{fullpage}
\usepackage{setspace}
\usepackage[backref=page]{hyperref}
\usepackage[nameinlink]{cleveref}
\usepackage{color}
\usepackage{wrapfig}
\usepackage{tikz}
\usetikzlibrary{decorations.pathreplacing}
\usepackage{algorithm}
\usepackage[noend]{algpseudocode}
\usepackage[framemethod=tikz]{mdframed}
\usepackage{xspace}
\usepackage{pgfplots}
\usepackage{framed}
\usepackage{thmtools}
\usepackage{thm-restate}
\usepackage{tabu}
\usepackage{fancyhdr}
\usepackage{booktabs,tabularx}
\pgfplotsset{compat=1.5}

\newtheorem{theorem}{Theorem}[section]
\newtheorem{corollary}[theorem]{Corollary}
\newtheorem{lemma}[theorem]{Lemma}
\newtheorem{proposition}[theorem]{Proposition}
\newtheorem{definition}[theorem]{Definition}
\newtheorem{remark}[theorem]{Remark}

\newtheorem{invariant}[theorem]{Invariant}
\newtheorem{observation}[theorem]{Observation}

\newenvironment{proofof}[1]{\begin{trivlist} \item {\bf Proof
#1:~~}}
  {\qed\end{trivlist}}

\newcommand{\namedref}[2]{\hyperref[#2]{#1~\ref*{#2}}}

\newcommand{\PPr}[1]{\ensuremath{\mathbf{Pr}\left[#1\right]}}

\newcommand{\Ex}[1]{\ensuremath{\mathbb{E}\left[#1\right]}}
\newcommand{\EEx}[2]{\ensuremath{\underset{#1}{\mathbb{E}}\left[#2\right]}}
\renewcommand{\O}[1]{\ensuremath{\mathcal{O}\left(#1\right)}}
\newcommand{\tO}[1]{\ensuremath{\tilde{\mathcal{O}}\left(#1\right)}}
\newcommand{\eps}{\varepsilon}

\def \calA    {\mdef{\mathcal{A}}}

\def \calC    {\mdef{\mathcal{C}}}
\def \calD    {\mdef{\mathcal{D}}}
\def \calE    {\mdef{\mathcal{E}}}

\def \calP    {\mdef{\mathcal{P}}}

\def \calU    {\mdef{\mathcal{U}}}

\def \calW    {\mdef{\mathcal{W}}}

\def \bA    {\mdef{\mathbf{A}}}
\def \bB    {\mdef{\mathbf{B}}}
\def \bC    {\mdef{\mathbf{C}}}

\def \bq    {\mdef{\mathbf{q}}}

\def \bu    {\mdef{\mathbf{u}}}
\def \bw    {\mdef{\mathbf{w}}}
\def \bv    {\mdef{\mathbf{v}}}
\def \bx    {\mdef{\mathbf{x}}}
\def \by    {\mdef{\mathbf{y}}}
\def \bz    {\mdef{\mathbf{z}}}

\newcommand{\mdef}[1]{{\ensuremath{#1}}\xspace}  

\DeclareMathOperator*{\argmin}{argmin}
\DeclareMathOperator*{\argmax}{argmax}
\DeclareMathOperator*{\polylog}{polylog}
\DeclareMathOperator*{\poly}{poly}
\DeclareMathOperator*{\Lap}{Lap}

\def \EstLevel    {\mdef{\mathsf{EstLevel}}}
\def \MaintainIter    {\mdef{\mathsf{MaintainIter}}}
\def \Accept    {\mdef{\mathsf{Accept}}}
\def \High    {\mdef{\mathsf{High}}}
\def \Low    {\mdef{\mathsf{Low}}}

\newcommand{\ignore}[1]{}

\newif\ifnotes\notestrue 
\ifnotes
\newcommand{\samson}[1]{\textcolor{blue}{{\bf (Samson:} {#1}{\bf ) }} \marginpar{\tiny\bf
             \begin{minipage}[t]{0.5in}
               \raggedright S:
            \end{minipage}}}
\newcommand{\david}[1]{\textcolor{purple}{{\bf (David:} {#1}{\bf ) }} \marginpar{\tiny\bf
             \begin{minipage}[t]{0.5in}
               \raggedright D:
            \end{minipage}}} 
\else
\newcommand{\samson}[1]{}
\newcommand{\david}[1]{}
\fi

\makeatletter
\renewcommand*{\@fnsymbol}[1]{\textcolor{mahogany}{\ensuremath{\ifcase#1\or *\or \dagger\or \ddagger\or
 \mathsection\or \triangledown\or \mathparagraph\or \|\or **\or \dagger\dagger
   \or \ddagger\ddagger \else\@ctrerr\fi}}}
\makeatother

\providecommand{\email}[1]{\href{mailto:#1}{\nolinkurl{#1}\xspace}}

\definecolor{mahogany}{rgb}{0.75, 0.25, 0.0}
\definecolor{darkblue}{rgb}{0.0, 0.0, 0.55}
\definecolor{darkpastelgreen}{rgb}{0.01, 0.75, 0.24}
\definecolor{darkgreen}{rgb}{0.0, 0.2, 0.13}
\definecolor{darkgoldenrod}{rgb}{0.72, 0.53, 0.04}
\definecolor{darkred}{rgb}{0.55, 0.0, 0.0}
\definecolor{forestgreenweb}{rgb}{0.13, 0.55, 0.13}
\definecolor{greencss}{rgb}{0.0, 0.5, 0.0}
\definecolor{bleudefrance}{rgb}{0.19, 0.55, 0.91}
\definecolor{darkpastelpurple}{rgb}{0.59, 0.44, 0.84}
\definecolor{darkcerulean}{rgb}{0.03, 0.27, 0.49}
\definecolor{royalpurple}{rgb}{0.47, 0.32, 0.66}

\hypersetup{
     colorlinks   = true,
     citecolor    = mahogany,
	linkcolor	  = darkpastelgreen,
     urlcolor   = mahogany
}

\fancypagestyle{pg}
{
\lhead{}
\rhead{}
\cfoot{--\ \thepage\ --}

}

\AtBeginDocument{%
  \DeclareFontShape{T1}{lmr}{m}{scit}{<->ssub*lmr/m/scsl}{}%
}

\begin{document}

\allowdisplaybreaks

\author{
Elena Gribelyuk
\thanks{Princeton University and Google Research. 
\email{eg5539@princeton.edu}. 
Supported in part by an NSF CAREER award CCF-2339942.
} 
\and
Honghao Lin
\thanks{Google Research. 
\email{honghaol3010@gmail.com}. 
} 
\and
Xin Lyu
\thanks{ByteDance. 
\email{lyuxin1999@gmail.com}.
} 
\and
Mahbod Majid
\thanks{MIT. 
\email{mahbod@mit.edu}. 
Supported by the 2026 Apple Scholars in AI/ML PhD fellowship.
}
\and
David P. Woodruff
\thanks{Carnegie Mellon University and Google Research. 
\email{dwoodruf@andrew.cmu.edu}. 
Supported in part by Office of Naval Research award number N000142112647 and a Simons Investigator Award. 
} 
\and
Huacheng Yu
\thanks{Princeton University. 
\email{hy2@cs.princeton.edu}.
Supported in part by an NSF CAREER award CCF-2339942.
} 
\and
Samson Zhou
\thanks{Texas A\&M University. \email{samsonzhou@gmail.com}. 
Supported in part by NSF CCF-2335411.}
}

\title{Adaptive Data Analysis and Adversarial Robustness on Insertion-Deletion Streams}

\begin{titlepage}
\maketitle
\thispagestyle{empty}

\begin{abstract}
Classical streaming algorithms and statistical mechanisms frequently fail when future inputs or queries depend adaptively on prior outputs. In this work, we resolve central open questions in both adaptive data analysis (ADA) and adversarially robust turnstile streaming by introducing a unified hierarchical estimator-corrector-learner framework. For adaptive data analysis over an unrestricted domain, we design a randomized mechanism that answers $k$ adaptively chosen statistical queries to within additive error $\alpha$ using $\tilde{\mathcal{O}}(\frac{\sqrt{k}}{\alpha} + \frac{1}{\alpha^2})$ samples, nearly matching the lower bounds of Hardt and Ullman [FOCS 2014] and Steinke and Ullman [COLT 2015] while improving upon the classic $\tilde{\mathcal{O}}(\frac{\sqrt{k}}{\alpha^2})$ bound of Dwork et al.\ [STOC 2015] and Bassily et al.\ [STOC 2016]. For known finite domains $\mathcal{U}$, we combine our mechanism with private multiplicative weights to improve the sample complexity to $\tilde{\mathcal{O}}\left(\frac{1}{\alpha^2} + \min\left(\frac{\sqrt{k}}{\alpha}, \frac{\sqrt{\log |\mathcal{U}|}}{\alpha^2}\right)\right)$. Building on our ADA mechanism, we construct a fixed linear sketch with a stateful decoder that answers $\tilde{\mathcal{O}}(\frac{1}{\varepsilon^2})$ adaptive $F_2$ moment estimation queries to within a $(1 \pm \varepsilon)$-multiplicative factor using $\tilde{\mathcal{O}}(\frac{1}{\varepsilon^2})$ bits of space. Finally, by embedding this sketch into a binary-tree recursion that preserves child states across parent updates via multiplicative rescaling, we obtain the first nearly space-optimal adversarially robust algorithm for $F_2$ estimation on insertion-deletion streams. Our algorithm uses $\frac{1}{\varepsilon^2} \cdot \operatorname{polylog}(n)$ bits of space, improving upon the $\tilde{\mathcal{O}}(\frac{1}{\varepsilon^4})$ bound of Gribelyuk et al.\ [STOC 2026] and matching the insertion-only bound of Woodruff and Zhou [FOCS 2021]. This demonstrates that no asymptotic space overhead in $\varepsilon$ is required to achieve adversarial robustness in the turnstile model. Our results have been verified in Lean.
\end{abstract}

\end{titlepage}

\section{Introduction}

Over the past three decades, the streaming model has emerged as a central paradigm for studying algorithms which process massive amounts of data. Specifically, updates to an underlying dataset arrive sequentially, and the task of the algorithm is to maintain a compact summary (or a \textit{sketch}) which supports dynamic updates and allows the user to recover various statistics about the dataset. Typically, the algorithm is only given a single pass over the data stream and must use space which is sublinear in both the universe size $n$ and stream length $m$. As the streaming model effectively captures data-processing pipelines in practical systems, there have been numerous works aimed at achieving optimal space streaming algorithms for a variety of problems.

It is a core assumption in the design and analysis of randomized algorithms that the input sequence is fixed in advance and independent of the internal parameters of the algorithm. However, in many applications, the intermediate responses of the algorithm may influence the distribution of future inputs to the system. For example, optimization procedures update intermediate iterates based on prior estimates, future queries to a database often depend on previous query outputs, and investors respond to changes in financial markets, which in turn shape future transactions. A special case of this paradigm is studied in adaptive data analysis, where a randomized mechanism is given a sample of i.i.d. points from an unknown distribution before the query sequence arrives, and the task of the algorithm is to return an accurate estimate for the population mean of each query $f_t$ on the underlying distribution, where queries may depend adaptively on prior estimates. While a common technique in the non-adaptive setting would be to return the empirical average of the sample points, an adaptive analyst can select queries which overfit to the sample, causing a significant deviation from the population mean. Likewise, the correctness analyses of classical streaming algorithms are completely invalidated when the update sequence may depend on the internal random bits (or sample points) used by the algorithm, necessitating the development of new approaches to reason about and ensure correctness. To this end, our work focuses on both settings simultaneously, developing techniques which are broadly applicable to both adaptive data analysis and adversarial streaming. Next, we describe each problem setting and motivation separately, beginning with the adversarially robust streaming model. 

\paragraph{Adversarially robust streaming.} 
Motivated by known limitations, in recent years there has been a large body of work exploring new algorithmic and lower bound techniques for algorithms in the adaptive setting \cite{Ben-EliezerJWY22, KaplanMNS21, HassidimKMMS20, CohenNSS23, ChakrabartiGS22, LinSWXZ25, CohenGNS26}. This paradigm has previously received significant attention in other settings, such as statistical queries in adaptive data analysis \cite{HardtU14, SteinkeU15, DworkFHPRR15, DworkFHPRR15b, DinurSWZ23}, dynamic graph algorithms and data structures \cite{BernsteinBGNSS022, AlokhinaB24, Wajc20, CherapanamjeriN20, AssadiYazdanyar26}, and adversarially robust machine learning \cite{AthalyeEIK18, GoodfellowSS14, HuangPGDA17}. Concretely, the data stream consists of $m$ updates $u_t = (a_t, \Delta_t)$ to the frequency vector $\bx \in \mathbb{Z}^n$, where $a_t \in [n]$ is an index and $|\Delta_t| \leq \poly(n)$ denotes the increment or decrement to index $a_t$. The update sequence induces an intermediate frequency vector $\bx^{(t)}$, where each entry is given by $\bx_i^{(t)} = \sum_{s \leq t: a_s = i} \Delta_s$. Then, at each time $t \in [m]$, the algorithm should return an accurate estimate of a given statistic $f(\bx^{(t)})$, even when updates may be selected adaptively based on prior algorithm outputs. 

\begin{definition}[\cite{Ben-EliezerJWY22}]
    Let $f: \mathbb{R}^n \rightarrow \mathbb{R}$ be a fixed function, and let $\bx^{(t)}$ denote the frequency vector at time $t \leq m$. A streaming algorithm $\calA$ is adversarially robust if, for any accuracy parameter $\eps \in (0,1)$, failure probability $\delta \in (0,1)$, and an adaptively chosen stream length $m= \poly(n)$, the algorithm $\calA$ returns a sequence of outputs $\{Z_i\}_{i \in [m]}$ satisfying 
    
    \[\Pr\left[\exists \ t \in [m] : |Z_t - f(\bx^{(t)})| > \eps f(\bx^{(t)})\right] \leq \delta.\]
\end{definition}

Alternatively, the adversarial streaming setting is represented by the following two-player game between the bounded-space algorithm and a computationally-unbounded adaptive adversary. 

\begin{enumerate}
    \item At each time $t \in [m]$, the adversary chooses an update $u_t$ to insert into the stream; $u_t$ may depend arbitrarily on the transcript of previous updates $\{u_{i}\}_{i < t}$ and algorithm responses $\{Z_{i}\}_{i < t}$.
    \item The algorithm receives update $u_t$, updates its internal state, and returns the current estimate $Z_t$ for $f(\bx^{(t)})$.
    \item The adversary records the algorithm's output and proceeds to the next round.
\end{enumerate}

The last few years have seen significant progress in the design of new algorithmic techniques for robustness, as well as strong lower bounds against common classes of streaming algorithms. For the special case that the stream contains only insertions of universe elements, the initial works of \cite{Ben-EliezerJWY22, HassidimKMMS20} presented general frameworks to ``robustify'' classical streaming algorithms for a large class of statistical estimation problems with only a small $\poly\left(\frac{1}{\eps}, \log n \right)$ overhead in the space complexity. This was subsequently improved by \cite{WoodruffZ21} to obtain tight bounds with respect to the accuracy parameter; for instance, their algorithm for $F_2$ moment estimation \footnote{The $F_p$ moment of the frequency vector $\bx \in \mathbb{R}^n$ is defined as $F_p := \|\bx\|_p^p = \sum_{i \in [n]} |x_i|^p$. For $p = 0$, the $F_0$ moment is  $\|\bx\|_0 := \left |\{i \in [n] : x_i \not = 0\} \right |$.} uses $\tO{\frac{1}{\eps^2}}$ bits of space, nearly matching the space complexity of $F_2$ in the classical streaming model \cite{AlonMS99, KaneNW10a}. From the other perspective, some works have uncovered space complexity separations between adaptive and non-adaptive streaming \cite{ChakrabartiGS22, KaplanMNS21}; in particular, \cite{KaplanMNS21} gave an exponential separation for the ``streaming'' version of the adaptive data analysis problem, formally connecting these two areas.

In the general case where the stream may contain both insertions and deletions, the situation was much more challenging. As space-optimal turnstile streaming algorithms are equivalent to linear sketches \cite{LiNW14, JiangLY26}, many works have focused on designing efficient adaptive attacks against linear-sketching based algorithms \cite{HardtW13, CohenNSSS24, GribelyukLWYZ24, AhmadianCS25, GribelyukLWYZ25}, as well as for specific sketches \cite{CohenNSS23, AhmadianCohen2024}. In fact, these adaptive attacks against algorithms which use \textit{persistent randomness} are conceptually equivalent to lower bounds for adaptive data analysis and interactive fingerprinting codes \cite{HardtU14, SteinkeU15}. For counting distinct elements, \cite{GribelyukLWYZ24} used fingerprinting-based techniques to show that any streaming algorithm which remains accurate under adaptive insertions and deletions and uses linear sketching-based methods must store at least $\Omega(n^{o(1)})$ bits of space. Similarly, \cite{GribelyukLWYZ25} showed that an adaptive adversary can essentially learn the underlying sketch matrix used by an $F_p$ estimation linear sketch after $\poly(n)$ adaptive queries, implying an $\Omega(n)$ space lower bound. Meanwhile, all known robust algorithms for $F_p$ estimation required space polynomial in the stream length $m$ \cite{Ben-EliezerJWY22, HassidimKMMS20, Ben-EliezerEO22}. Very recently, several works \cite{GribelyukLWYZ26, GribelyukLWYZ26b} circumvented this $\Omega(n)$ barrier for linear sketches and designed the first robust algorithms for $F_p$ estimation for all $p \in [0,2]$ which store only $\poly\left(\frac{1}{\eps}, \log n\right)$ bits of space. To overcome prior lower bounds, these algorithms are crucially non-linear and introduce new random bits during the execution of the algorithm. 

\paragraph{$F_2$ moment estimation.}
In this work, we return to the $F_2$ moment estimation problem in the adversarial streaming model. The seminal work of \cite{AlonMS99} gave an  $\O{\frac{1}{\eps^2} \log n}$ space algorithm for $F_2$, which was later shown to be exactly optimal \cite{KaneNW10a, BravermanZ25}. Since then, the $F_2$ estimation problem has become particularly significant in both theory and practice. Theoretical techniques developed for $F_2$ estimation have inspired efficient streaming algorithms for a host of other estimation tasks, and
streaming lower bounds for $F_p$ estimation were the first to be proved via reductions from communication complexity \cite{AlonMS99, IndykWoodruff03, Woodruff04}, sparking a beautiful line of works which has further explored the connections between these models. On the practical side, the $F_2$ moment roughly represents the \textit{repeat rate} of the input stream, and algorithms for $F_2$ are often used in network traffic monitoring \cite{KrishnamurthySZC03} and database query optimization \cite{AlonGMS99}. 

In the adaptive setting, the space complexity of $F_2$ estimation is far from being well understood. Note that for insertion-only streams, \cite{WoodruffZ21} gave an $\tO{\frac{1}{\eps^2}}$ space adversarially robust algorithm for $F_2$, nearly matching the classical streaming lower bound of $\Omega\left(\frac{1}{\eps^2} \log n\right)$. While the recent robust $F_2$ algorithm \cite{GribelyukLWYZ26} achieves $\poly\left(\frac{1}{\eps}, \log n \right)$ space, the dependence on the accuracy parameter in the space complexity scales as $\tO{\frac{1}{\eps^4}}$. To this end, we study the following central question:

\begin{quote}
   \em{Is there a separation in the space complexity of $F_2$ moment estimation in the adversarially robust and classical turnstile streaming models?}
\end{quote}

To address this question, we first focus on a question of independent interest in adaptive data analysis, which we describe next.

\paragraph{Adaptive data analysis.}
Motivated by statistical analyses in natural sciences, an earlier line of work has explored the \textit{adaptive} data analysis problem, where a finite sample is drawn from an unknown distribution $\calD$, and several analyses are then performed to uncover properties of the underlying distribution $\calD$. In this setting, the outcome of an analysis is viewed as statistically significant if the outcome is unlikely to occur due to just random chance. However, practical data analysis processes are often inherently adaptive: analysts may elect to run certain tests after inspecting the sample set itself, thus increasing the chance of overfitting and false discovery \cite{HardtU14, SteinkeU15, DworkFHPRR15}. Indeed, this has been identified as a major concern in the statistics community over the past few years \cite{GelmanLoken14}. 

To this end, adaptive data analysis studies a theoretical formulation of this problem, where the mechanism is given an i.i.d. sample $X_1, \ldots, X_n \sim \calD$ from an unknown distribution $\calD$ with support $\calU$, and the task of the mechanism is to answer adaptive statistical queries about $\calD$. Formally, at each time $t \in [k]$, a data analyst provides a statistical query $f_t : \calU \rightarrow [0,1]$, and the mechanism should return an accurate estimate $a_t$ for the population mean $\mathbb{E}_{X \sim \calD} [f_t(X)]$ with high probability, such that $\left |a_t - \mathbb{E}_{X \sim \calD} f_t(X) \right | \leq \alpha$ for some accuracy parameter $\alpha \in (0,1)$. Importantly, each query $f_t$ requested by the analyst may depend on all prior mechanism estimates $\{a_i\}_{i < t}$ and queries $\{f_i\}_{i < t}$. 
For non-adaptive queries, the standard approach is to take the empirical average of $\tO{\frac{1}{\alpha^2}\log k}$ samples, which suffices by concentration and a union bound. 
For adaptive queries, this argument breaks down, since an analyst can gradually select queries that exploit the samples.  Note that using a fresh sample block for each query restores independence, but requires $\tO{\frac{k}{\alpha^2}}$ samples. 

Over the past decade, there has been substantial progress on designing improved mechanisms and sample lower bounds for adaptive data analysis in various settings \cite{HardtU14, DworkFHPRR15, BassilyNSSSU16, BassilyF16, FeldmanS18, NissimST23, RapoportCS25, LyuT25}. Known mechanisms broadly fall into one of two categories, which are distinguished by whether the sample complexity depends on the support size of the distribution $\calD$. In the unrestricted domain setting, the differential privacy approach of~\cite{DworkFHPRR15}, with the improved generalization analysis of~\cite{BassilyNSSSU21}, requires $\tO{\frac{\sqrt{k}}{\alpha^2}}$ samples. 
To see the dependence on $\alpha$, consider adding noise to each empirical answer. 
With $n$ samples and total privacy parameter $\eps$, composition gives noise of magnitude approximately $\frac{\sqrt{k}}{n\eps}$. 
To ensure that the empirical answers remain within additive error $\O{\alpha}$ of the population answer $\mathbb{E}_{X \sim \calD} [f_t(X)]$, the standard generalization bound requires setting the total privacy budget to $\eps=\O{\alpha}$. 
Requiring the noise to have magnitude $\O{\alpha}$ then gives $n=\tO{\frac{\sqrt{k}}{\alpha^2}}$. Alternatively, in the setting that the unknown distribution $\calD$ has bounded support $|\calU|$, one approach uses private multiplicative weights~\cite{HardtR10,BassilyNSSSU21} to achieve sample complexity roughly $\tO{\frac{1}{\alpha^3}\sqrt{\log|\mathcal U|}\log k}$. 
This mechanism maintains a hypothesis distribution, checks its predictions privately, and updates it when a query exposes an error. Note that this logarithmic dependence on $k$ can give a smaller sample bound when the number of queries is sufficiently large. On the lower bound side, \cite{HardtU14, SteinkeU15} showed that any mechanism which can accurately estimate $k$ statistical queries must maintain $n \geq \tilde \Omega\left(\frac{\min\left(\sqrt{k},  \ \sqrt{\log |\calU|}\right)}{\alpha}\right)$ samples. As answering a single statistical query requires $\Omega\left(\frac{1}{\alpha^2}\right)$ samples, this implies a lower bound of $\tilde \Omega\left(\frac{\min\left(\sqrt{k},  \ \sqrt{\log |\calU|}\right)}{\alpha} + \frac{1}{\alpha^2} \right)$. A central question \cite{Steinke16, BassilyNSSSU21} has therefore been to resolve this gap in the sample complexity of adaptive data analysis for both settings:

\begin{quote}
   \em{Is there a statistically accurate mechanism for adaptive data analysis  which answers $k$ adaptive statistical queries using $\tO{\frac{\min\left(\sqrt{k},\ \sqrt{\log |\calU|}\right)}{\alpha} + \frac{1}{\alpha^2}}$ samples?}
\end{quote}

\subsection{Our Contributions}

In this work, we significantly advance our understanding of robust streaming and adaptive data analysis, obtaining optimal bounds for both of the central problems introduced above. For the unrestricted domain setting, we  design a mechanism which uses $\tO{\frac{\sqrt{k}}{\alpha} + \frac{1}{\alpha^2}}$ samples, nearly matching the best known lower bound up to logarithmic factors in $k$ \cite{HardtU14, SteinkeU15}.

\begin{theorem}[Adaptive data analysis]
\label{thm:contributions:ada}
Given an accuracy parameter $\alpha\in(0,1)$ and failure probability $\delta\in(0,1)$, there exists a randomized mechanism that answers $k$ adaptively chosen queries $f_t:\calU\to[0,1]$ to additive error $\alpha$ with respect to any unknown distribution $D$ on a finite domain $\calU$, simultaneously with probability at least $1-\delta$, using $\left(\frac{\sqrt{k}}{\alpha}+\frac{1}{\alpha^2}\right)\cdot\polylog\left(\frac{2k}{\alpha\delta}\right)$ independent samples from $D$.
\end{theorem}

On the path to obtain our nearly-optimal $F_2$ estimation algorithm in the adversarial streaming setting, we consider the adaptive query model, where the algorithm samples a sketch matrix $\bA \in \mathbb{R}^{r \times n}$ before the input arrives, and for each time $t \in [k]$ the algorithm must provide a $(1 \pm \eps)$-approximation for the $F_2$ moment of a query vector $\bx_t$, which may be chosen adaptively based on prior algorithm outputs. Importantly, the algorithm can only access the query $\bx_t$ by observing the sketch $\bA \bx_t$. This adaptive query model has previously been studied by \cite{AhmadianCS25, CohenStemmer26}, which showed that any linear sketch with $r$ rows can be broken using $\tO{r + \eps^2 \cdot r^2}$ adaptive queries. Specifically, when $r = \tO{\frac{1}{\eps^2}}$, their attack uses $\tO{\frac{1}{\eps^2}}$ adaptive queries.
In fact, our next corollary gives an algorithm which exactly matches the query complexity of their attack for the case that $r = \tO{\frac{1}{\eps^2}}$. As optimal space linear sketches use $\tO{\frac{1}{\eps^2}}$ bits of space, this exactly characterizes the power of $F_2$ linear sketches under adaptivity. 

\begin{corollary}[Adaptive $F_2$ queries]
\label{cor:contributions:toy}
Given an accuracy parameter $\eps\in(0,1)$ and failure probability $\delta\in(0,1)$, there exists a fixed linear sketch with a stateful decoder that answers $\tO{\frac{1}{\eps^2}}$ adaptively chosen $F_2$ queries on vectors in $\mathbb Z^n$ with polynomially bounded coordinates, simultaneously to within a factor of $(1+\eps)$ with probability at least $1-\delta$, using $\frac{1}{\eps^2}\cdot\polylog\left(n,\frac{1}{\eps},\frac{1}{\delta}\right)$ bits of space.
\end{corollary}

As an adversarially robust streaming algorithm must return an estimate of the $F_2$ moment at each time $t \in [m]$ for an input stream of length $m = \poly(n)$, \Cref{cor:contributions:toy} does not immediately imply any improvement over the previous $\tO{\frac{1}{\eps^4}}$ space algorithm of \cite{GribelyukLWYZ26}.  Our main insight is a way to carefully combine the sketch in \Cref{cor:contributions:toy} with an improved algorithmic framework for robust $F_2$ estimation which uses each linear sketch for at most $\tO{\frac{1}{\eps^2}}$ adaptive queries throughout the stream of length $m$. We design a nearly-optimal $\tO{\frac{1}{\eps^2}}$ space algorithm for $F_2$ in the adversarial streaming model under insertions and deletions. In particular, this shows that no losses in $\eps$ are necessary to achieve robustness up to polylogarithmic factors in $n$. 

\begin{theorem}[Adversarially robust $F_2$ streaming]
\label{thm:contributions:f2}
Given an accuracy parameter $\eps\in(0,1)$ and a turnstile stream of length $m=\poly(n)$ on a universe of size $n$, with polynomially bounded integer frequencies, there exists an adversarially robust algorithm that, with probability at least $1-\frac{1}{\poly(n)}$, outputs a $(1+\eps)$-approximation to the $F_2$ moment at all times and uses $\frac{1}{\eps^2}\cdot\polylog(n)$ bits of space.
\end{theorem}

Finally, by combining our improved mechanism for the unrestricted domain setting in \Cref{thm:contributions:ada} with the private multiplicative weights framework of \cite{BassilyNSSSU21}, we obtain the following improved mechanism for the bounded universe setting of adaptive data analysis.

\begin{theorem}[Adaptive data analysis on a bounded universe]
\label{thm:contributions:ada:bounded}
Given an accuracy parameter $\alpha\in(0,1)$ and failure probability $\delta\in(0,1)$, there exists a randomized mechanism that answers $k$ adaptively chosen queries $f_t:\calU\to[0,1]$ to additive error $\alpha$ with respect to any unknown distribution $D$ on a known finite domain $\calU$, simultaneously with probability at least $1-\delta$, using $\left(\frac{\sqrt{\log|\calU|}}{\alpha^2}\right)\cdot\polylog\left(\frac{2k}{\alpha\delta}\right)$ independent samples from $D$.
\end{theorem}

\subsection{Technical Overview}

\subsubsection{Adaptive Data Analysis}
\label{sec:ada-technical-overview}
Recall that in adaptive data analysis, we receive independent samples from an unknown distribution $D$ over a finite domain $\mathcal U$ and must answer $k$ adaptively chosen queries $f_t:\mathcal U\to[0,1]$. 
Writing $Df=\mathbb E_{X\sim D}[f(X)]$, our goal is to return answers $a_1,\ldots,a_k$ satisfying $|a_t-Df_t|\le\alpha$ for all $t$. 
We describe how to achieve this using $N=\tO{\frac{\sqrt{k}}{\alpha}+\frac{1}{\alpha^2}}$ samples. 
In particular, the sample complexity is $\tO{\frac{1}{\alpha^2}}$ when $k=\tO{\frac{1}{\alpha^2}}$. 

\paragraph{Overview.}
Our first observation is that a good estimate to the population answer need not be a good estimate to the empirical average. 
Instead, our approach uses differential privacy to show that a set of samples admits a set of ``reference weights'' whose weighted empirical averages accurately answer all the adaptive queries. 
Although the algorithm does not know these reference weights, it maintains candidate weights and uses successive corrections to guide them toward an accurate reference weighting. 
The existence of such a reference weighting allows us to upper bound the total number of corrections across the $k$ queries. 

We first explain this idea through a simplified estimator-corrector-learner framework: an estimator answers queries using its current weights, a corrector detects inaccurate predictions, and a learner updates the weights in response. 
We then explain why protecting a single corrector is too expensive, and replace it with a hierarchy in which intermediate blocks also learn. 
The main step in the analysis shows that corrections become progressively rarer toward the top of the hierarchy, allowing these blocks to maintain stronger privacy guarantees without requiring more samples.

\paragraph{Existence of reference weights, i.e., accurate answers by reweighting samples.}
Consider a block $S=(X_1,\ldots,X_r)$ of $r$ samples whose interaction transcript is $(\eps,\delta)$-private. 
We first show that for $r=\O{\frac{1}{\tau^2}\log k}$, there exists a near-uniform probability vector $u\in\mathbb{R}^r$ such that
\[0\le u_i\le\frac{8}{r},\qquad R(u):=\frac r2\sum_{i=1}^r\left(u_i-\frac{1}{r}\right)^2\le\eps^2,\qquad\left|\sum_{i=1}^r u_i f_t(X_i)-Df_t\right|\le\frac{\tau}{64}\]
for every query in the realized interaction. 
The vector $u$ consists of the reference weights and is fixed across the $k$ queries, although it may depend on the completed transcript. 
Although the algorithm does not know or compute this vector, it nevertheless certifies that a ``near-uniform'' weighting of the observed points can answer all the selected queries accurately, which will ultimately be used to upper bound the total number of corrections made by the learner.

To see why such reference weights should exist, reveal the completed transcript after the $k$ queries and then reveal the samples one at a time. 
Let $p_i$ be the conditional distribution of $X_i$ given the transcript and the previously revealed samples. 
Then, weighting a value $x$ by $D(x)/p_i(x)$ would undo the conditional bias. 
Pure differential privacy would ensure that these ratios lie in $[e^{-\eps},e^\eps]$, but result in a sub-optimal sample complexity. 
Instead, we use approximate differential privacy to give a better sample bound through advanced composition and then to upper bound the resulting bias in expectation, after clipping the ratios to the interval $[e^{-\eps},e^\eps]$. 

While this controls the expected bias, we also need the reweighting to be accurate with high probability, simultaneously for all $k$ queries. 
This can be shown using a standard martingale concentration inequality. 
Finally, we normalize the correction factors to obtain reference weights that sum to one.
Observe that clipping keeps every correction factor within $1\pm\O{\eps}$, so after normalization each reference weight differs from the uniform $\frac{1}{r}$ by at most $\O{\frac{\eps}{r}}$. 
Thus the total quadratic deviation from uniform is $R(u)\le\O{\eps^2}$ and can be sharpened appropriately. 

\paragraph{A simplified estimator-corrector-learner.}
To explain how to find useful weights without knowing $D$, we consider the following core components:
\begin{enumerate}
\item 
An \emph{estimator} maintains a probability vector $w$ on $S$ and outputs the prediction $\mu(f)=\sum_iw_if(X_i)$.
\item 
A \emph{corrector} checks whether this prediction differs ``substantially'' from $Df$. 
If so, it supplies a sign $q\in\{-1,+1\}$ indicating the direction in which the prediction should move.
\item 
A \emph{learner} uses this failed prediction to update $w$, so that the same sample can support future queries.
\end{enumerate}
The learner stores scores $G_i$, initially zero. 
On a correction it adds $qf(X_i)$ to each score, and chooses
\[w(G)=\mathop{\rm argmax}_{w\in W}\left(\frac{\tau}{4}\langle G,w\rangle-R(w)\right),\qquad W=\left\{w\ge0:\sum_iw_i=1,\ w_i\le\frac8r\right\}.\]
For example, when the prediction is too small, points with larger query values receive larger score increments. 
The quadratic penalty limits how aggressively the weights move, while the cap prevents a few sample points from carrying excessive mass.

Suppose the corrector detects that the estimator's prediction $\mu(f)=\langle w,f(S)\rangle$ is inaccurate, and returns the correction sign $q$. 
Since the reference prediction $\langle u,f(S)\rangle$ is accurate, the chosen accuracy margins give
\[q\left(\langle u,f(S)\rangle-\langle w,f(S)\rangle\right)\ge \frac{\tau}{2}.\]
Thus, the corrector identifies a query on which the estimator's current weights and the reference weights give substantially different predictions. 
The learner responds by updating $G\leftarrow G+qf(S)$ and recomputing its weights. 
To quantify the progress made by this correction, let
\[\Phi(G):=\max_{v\in W}\left(\frac{\tau}{4}\langle G,v\rangle-R(v)\right),\]
and consider the potential
\[\Psi(G):=\Phi(G)-\left(\frac{\tau}{4}\langle G,u\rangle-R(u)\right).\]
This potential is non-negative because the reference weights $u$ are feasible for the maximization defining $\Phi(G)$.  
Initially, $G=0$ and the learner uses uniform weights, so $\Psi(0)=R(u)\le\eps^2$.

When the learner updates the scores, the objective evaluated at
the fixed reference weights increases by
$\frac{\tau}{4}q\langle u,f(S)\rangle$.
If the estimator kept its old weights $w$, its objective would
increase by $\frac{\tau}{4}q\langle w,f(S)\rangle$.
The former increase exceeds the latter by at least $\frac{\tau^2}{8}$. 
The learner also recomputes the maximizing weights, which can further increase $\Phi$. 
However, the quadratic regularizer limits this additional increase to at most $\frac{\tau^2}{32}$, which implies that the potential changes by
\[\Psi(G+qf(S))-\Psi(G)\le-\frac{\tau}{4}q\langle u-w,f(S)\rangle+\frac{\tau^2}{32}\le -\frac{3\tau^2}{32}.\]
Thus, every error detected by the corrector leads to a learner update that decreases the potential by $\Omega(\tau^2)$.
Since the potential starts at most $\eps^2$ and remains nonnegative, at most $\O{\frac{\eps^2}{\tau^2}}$ corrections can occur in this simplified setting.

\paragraph{Why one corrector is not enough.} 
The remaining difficulty is to implement a corrector that can be trusted over the entirety of the adaptive interaction. 
A natural attempt uses an independent hidden ``holdout'' block, keeps its weights uniform, and privately checks whether the estimator disagrees with its empirical average.  
\textsc{AboveThreshold} allows many unsuccessful checks without paying separately for each one. 
However, the holdout still needs enough privacy to remain accurate throughout the corrections.

For intuition, take $\tau=\Theta(\alpha)$ and constant privacy for the learned estimator. 
The previous argument shows that the number of corrections is at most $L=\O{\frac{1}{\alpha^2}}$, and so it suffices to have a holdout protected with privacy $\O{\alpha}$ and noise with magnitude $\O{\alpha}$, which gives overall sample complexity $\tO{\frac{\sqrt{L}}{\alpha\cdot\alpha}}$. 
Thus a single holdout block results in an expensive number of samples to ensure correctness. 
Our solution is to let the corrector learn as well, and protect it using another block with fewer corrections.

\paragraph{A hierarchy of estimators, correctors, and learners.}
We use independent sample blocks $S_1,\ldots,S_p$, where $p=\O{\log\frac1\alpha}$, and set the error threshold at each level to $\tau=\frac{\alpha}{4p}$. 
Each block $j<p$ maintains scores $G_j$, weights $w_j$, and a prediction $\mu_j(f)=\langle w_j,f(S_j)\rangle$. 
Each intermediate block checks the block below it and is itself checked by the block above it. 
The final block is a uniformly weighted \emph{anchor}, and the output to each adaptive query is given by the first block. 

For each query, we compare neighboring predictions using private \textsc{AboveThreshold} tests. 
If a discrepancy of at least $\tau$ is reported between levels $j$ and $j+1$, the learner adjusts the scores $G_j$ and $G_{j+1}$, which causes the resulting predictions to move toward each other. 
Since the disagreement does not identify which prediction is inaccurate, both learned blocks are updated, with the upper block receiving a smaller adjustment, though the anchor is never changed. 
After each correction, we recompute the affected weights $w_j$ and restart the scan from the first block $S_1$, since changing one prediction may have invalidated an earlier comparison. 
The scores $G_j$ and weights $w_j$ persist across corrections and queries, and once a complete scan finds no disagreement, the algorithm releases $\mu_1(f)$, i.e., the prediction by the first block, along with some added noise to maintain privacy. 

\paragraph{Progress and privacy across the hierarchy.}
To avoid a larger number of necessary samples, the key property is that blocks with stronger privacy guarantees undergo fewer corrections. 
Recall that in the simplified setting, a learner with privacy budget $\eps$ has reference weights whose quadratic distance from uniform is at most $\eps^2$. 
This bounds the initial potential, while each correction decreases it by $\Omega(\tau^2)$, giving at most $\O{\frac{\eps^2}{\tau^2}}$ corrections.
To extend this relationship to the hierarchy, we analyze progress jointly: although neighboring predictions may both be inaccurate, their reference predictions approximate the same population answer. 
The coupled updates therefore decrease a shared potential, even when one block's update is unhelpful in isolation. 
This yields correction bounds decreasing by a factor of four as privacy budgets decrease by a factor of two per level.

Moreover, the regularized predictor has sensitivity $\O{\frac{1}{n_j}}$, while \textsc{AboveThreshold} makes the privacy cost depend on the number of corrections. 
For an intermediate block, we restrict the number of corrections to be at most $B_j=\O{4^{-(j-1)}/\tau^2}$ and privacy budget $\eps_j=2^{-(j-1)}/\sqrt p$. 
Then by advanced composition, the necessary sample complexity is $n_j=\tO{\frac{\sqrt{B_j}}{\tau\eps_j}}=\tO{1/\alpha^2}\cdot\polylog(k)$, since $\tau=\alpha/(4p)$ and $p=\O{\log(1/\alpha)}$. 
We choose the depth $p$ so that the anchor only has a constant number of corrections, so the same sample bound gives privacy $\eps_p=\Theta(\tau)$ and error $\O{\tau}$ for its uniform averages. 
The first block additionally releases $k$ numerical answers; its privacy budget $\frac{1}{\sqrt{p}}$ gives an additional sample cost of $\tO{\sqrt{k}/\alpha}$.

We remark that there is a slight concern about circular reasoning, since privacy depends on the number of corrections and vice versa. 
We circumvent this issue by first imposing deterministic limits on the number of corrections and establishing privacy for the resulting mechanism. 
We then obtain the reference weights and use the progress analysis to show that these limits are never reached with high probability.

Once a scan succeeds, adjacent predictions differ by $\O{\tau}$, and the anchor has population error $\O{\tau}$. 
Telescoping across the hierarchy gives error $\O{p\tau}$ for the first block; the chosen constants ensure error at most $\alpha$, including output noise and rounding. 
The correction bound guarantees that every query finishes. 
Summing over the logarithmic number of blocks gives $\tO{\frac{\sqrt{k}}{\alpha}+\frac{1}{\alpha^2}}$ samples, as desired. 

\subsubsection{A Toy Problem: Adaptive \texorpdfstring{$F_2$}{F2} Queries}
\label{sec:f2-toy-overview}
We first isolate a toy problem connecting our ADA result to adversarially robust streaming.
Starting from $\bx=\mathbf{0}\in\mathbb Z^n$, we process turnstile updates and answer only $K=\O{\frac{1}{\eps^2}}$ queries for $F_2(\bx)=\|\bx\|_2^2$. 
Both the updates and the query times may depend on previously reported answers, and there can be up to $\poly(n)$ updates between queries, so that $\|x\|_\infty\le\poly(n)$. 
The adversary sees the previous algorithm outputs but not its internal state. 
We show that to answer all $K$ queries within a multiplicative factor of $(1+\eps)$, simultaneously with probability at least $1-\delta$, it suffices to use $\frac{1}{\eps^2}\cdot\polylog\left(n,\frac{1}{\delta}\right)$ bits of space. 

A standard approach for this toy problem would be to reserve a separate accurate sketch for each query, maintaining all of them from the start of the stream. 
Robustness is immediate since each sketch is only used once, but with any standard $F_2$ streaming algorithm such as AMS~\cite{AlonMS99}, this requires $\tO{\frac{K}{\eps^2}}\cdot\polylog(n)=\tO{\frac{1}{\eps^4}}\cdot\polylog(n)$ space. 
The usual differential privacy approach improves the dependence to $\tO{\frac{\sqrt{K}}{\eps^2}}=\tO{\frac{1}{\eps^3}}\cdot\polylog(n)$. 
Our ADA mechanism removes the remaining $\frac{1}{\eps}$ loss by reusing the same measurements across adaptive queries.

\paragraph{From adaptive statistical queries to norm estimates.}
For a random sign vector $s\in\{\pm1\}^n$, we have $\EEx{s}{\langle s,v\rangle^2}=\|v\|_2^2$. 
Thus, we can view independently sampled sign rows as the samples of our ADA mechanism, and a vector $v$ as specifying a query on these rows. 
The streaming algorithm maintains the linear measurements $\langle b,v\rangle$, which suffice to evaluate the query without storing $v$. 
The difficulty is that squared measurements are unbounded, whereas the ADA mechanism requires bounded queries $f:\{\pm 1\}^n\to[0,1]$. 
To this end, we use a constant-factor approximation $C$ of $\|v\|_2^2$ to find a good scale. 
Namely, we set $f_{v,c}(b)=\min\left(\frac{\langle b,v\rangle^2}{Lc},1\right)$, where $L=\O{\log\frac{1}{\eps}}$. 
Moment bounds for random signs show that this clipping only introduces $\O{\eps\|v\|_2^2}$ bias. 
However, the benefit is that additive error $\O{\frac{\eps}{L}}$ for this query translates to $(1+\eps)$-multiplicative error for the norm. 
To acquire the constant-factor approximation $c$, we maintain a separate sketch for each query. 
This uses $K\cdot\polylog(n)$ space, so that along with the $\tO{\frac{\sqrt{K}}{\eps}+\frac{1}{\eps^2}}$ samples from our ADA mechanism, the total space is 
\[\tO{K+\frac{\sqrt K}{\eps}+\frac{1}{\eps^2}}\cdot\polylog(n)=\frac{1}{\eps^2}\cdot\polylog(n).\]

\subsubsection{Adversarially Robust \texorpdfstring{$F_2$}{F2} Streaming}
\label{sec:f2-technical-overview}
Unfortunately, the full streaming problem requires an answer after \emph{every} update, i.e., a stream of length $m$ has $m$ adaptive queries, which can be much larger than $\frac{1}{\eps^2}$. 
Applying the toy problem directly with $K=m$ would introduce a polynomial dependence on the stream length. 
We also cannot simply restart the problem after each batch of $\frac{1}{\eps^2}$ queries, since a newly formed sketch does not have access to the previous stream updates. 
We instead provide a reduction to the toy problem using the estimator-corrector-learner framework of \cite{GribelyukLWYZ26}. 

\paragraph{The estimator, corrector, and learner.}
Suppose that $\bx=\bz+\bq$, where $\bz$ is the frequency vector at some earlier time and $\bq$ contains the subsequent updates. 
An older sketch $\bA$ has seen the prefix and maintains both $\bA\bz$ and $\bA\bx$. An independent sketch $\bB$, initialized at that earlier time, can maintain $\bB\bq$ from zero, but cannot recover $\bB\bz$. 
The goal is to use the newer sketch to answer most queries while requesting few numerical estimates from the older sketch. 
To do so, the learner maintains an iterate $\bv$, initially zero, representing its current approximation to $\bz$.
The three components operate as follows:
\begin{enumerate}
\item 
The \emph{estimator} submits an approximation of $\|\bz-\bv\|_2^2+\|\bv+\bq\|_2^2$ as an estimate of $\|\bx\|_2^2$. 
The first term is estimated using $\bA$ and cached until $\bv$ changes. 
The second is obtained from the newer measurements $\bB(\bv+\bq)$, using the recursion described below. 
\item 
The \emph{corrector} uses $\bA\bx$ to check the proposed output by the estimator. 
If the discrepancy is small, the estimator's answer is released. 
Otherwise, the corrector supplies an accurate answer for the current time-step and notifies the learner that the proposed estimate was too large or too small.
\item 
The \emph{learner} uses an incorrect notification to update $\bv$ and refresh the cached estimate of $\|\bz - \bv\|_2^2$. 
The updated iterate is retained for subsequent queries.
\end{enumerate}
All vectors are represented through their linear measurements. 
In particular, we maintain $\bA\bv$ and $\bB\bv$, rather than storing $\bv$ explicitly.

\paragraph{Why an inaccurate estimate permits learning.}
The estimator omits exactly one cross term,
\[\|\bx\|_2^2-\left(\|\bz-\bv\|_2^2+\|\bv+\bq\|_2^2\right)=2\langle\bz-\bv,\bv+\bq\rangle,\]
and so if the cross-term is small, the estimator is already accurate. 
Otherwise, $\bv+\bq$ gives a direction correlated with the remaining error $\bz-\bv$.
Let $\eta$ be the local accuracy parameter, chosen below to accommodate the recursion depth. 
The threshold for the corrector is chosen so that a rejection certifies a cross term of magnitude $\Omega(\eta\|\bz-\bv\|_2\|\bv+\bq\|_2)$ with the indicated sign. 
The learner then takes a step
\[\bv'=\bv+\lambda(\bv+\bq),\]
where $\lambda$ has this sign and magnitude proportional to $\eta\|\bz-\bv\|_2/\|\bv+\bq\|_2$, computed from the available estimates. 
Expanding the squared residual shows that a sufficiently small step gives
\[\|\bz-\bv'\|_2^2\le \left(1-\Omega(\eta^2)\right)\|\bz-\bv\|_2^2.\]
Thus every correction removes a fixed fraction of the remaining squared error, after which we can conclude that the learner only requires $L:=\O{\frac{1}{\eta^2}\log n}$ corrections.  

\paragraph{The cost of restarting recursive estimators.}
Estimating $\|\bv+\bq\|_2^2$ is itself a streaming norm-estimation problem, so it can be handled by implementing the estimator-corrector-learner framework recursively, i.e. by approximating $\|\bv + \bq\|_2^2$ using the estimator $\|\bv + \bq_1 - \bu\|_2^2 + \|\bu + \bq_2\|_2^2$, where $\bq = \bq_1 + \bq_2$ and $\bu$ is the iterate for this estimation instance. However, the difficulty is that the child node's input changes whenever its parent updates $\bv$. 
The construction of~\cite{GribelyukLWYZ26} handles such changes by starting a new child node each time that the parent's iterate changes. Its recursion uses a branching factor $B\ge L$, and each corrector serves $\O{B}$ possible targets for the learner, each requiring $\O{L}$ possible corrections, leading to $BL=\frac{1}{\eps^4}\cdot\polylog(n)$ corrections. 
Applying the differential privacy robustification technique of~\cite{HassidimKMMS20} in tandem with a standard $F_2$ sketching algorithm such as AMS gives a total space of $\sqrt{BL}\cdot\frac{1}{\eta^2}\cdot\polylog(n)=\frac{1}{\eta^4}\cdot\polylog(n)$ bits. 
We emphasize that replacing the statistical mechanism alone would still leave this larger request budget in place, leading to sub-optimal space.

\paragraph{Preserving the child when the parent learns.}
Our first improvement is to retain the recursion in the child node when a parent node's iterate is updated.
In fact, the update rule above has the useful property that
\[\bv'+\bq=(1+\lambda)(\bv+\bq).\]
Thus the child's entire current input $\bv'+ \bq_1 + \bq_2$ is multiplied by one known positive factor. 
We apply this factor to its prefix, suffix, and learned iterate, and its square to the cached residual estimate of $\|(\bv + \bq_1) - \bu\|_2^2$.
These changes can be performed directly on the stored measurements and propagated to the active descendants in the recursion. 
After dividing out the accumulated rescaling from its ancestors, the child is still learning the same fixed prefix. In particular, this means that the progress it has already made toward learning $(\bv + \bq_1)$ is preserved, and only updates to its own iterate $\bu$ will affect the total number of adaptive queries which the sketch must handle. Note that these rescalings act on these virtual states; future physical updates retain their original values.

We can consequently use a binary tree of time intervals fixed before the stream begins, with depth $H=\O{\log(m+1)}$. 
Each block has one child for each half of its interval, and corrections never create additional child nodes. 
A block uses its parent's sketch to check its incoming vector, and initializes its own sketch to track $\bv+\bq$, which starts at zero. 
This sketch then serves as the corrector for its two children. 
Hence, each corrector handles only $\O{L}$ corrections in total. 
Since only one root-to-leaf path is active, and the sparse vectors at the leaves can be evaluated explicitly; by applying the previous differential privacy robustness approach, this reduces the total space to $\frac{\sqrt{L}}{\eta^2}\cdot\polylog(n)=\frac{1}{\eps^3}\cdot\polylog(n)$. 

\paragraph{Troubleshooting the toy problem: precision and space.}
Our second improvement replaces this differential privacy-based robustness framework with the ADA mechanism for the toy problem.
Each sketch now needs only $\tO{\frac{1}{\eta^2}}$ corrections, which is precisely the regime in which the solution to the toy problem uses $\frac{1}{\eta^2}\cdot\polylog(n)$ space. 
There is one additional requirement: the corrector checks a proposal after every update, and accepted checks also reveal information about its sketch. 
We therefore use the verification extension of the ADA mechanism, so that persistent \textsc{AboveThreshold} phases allow many accepted checks while charging privacy to rejections and numerical releases. 
Since each correction requires a constant number of numerical estimates, then this verification extension uses $\frac{1}{\eta^2}\cdot\polylog(n)$ sketching dimension, as desired. 

Finally, to translate to bit complexity, we round all maintained vectors to $\frac{1}{\poly(n)}$ precision after every step. However, we must then control rounding errors under the repeated rescalings which appear in the update rule $\bv' + \bq = (1+\lambda) (\bv + \bq)$. Note that each level will experience $\tO{\frac{1}{\eta^2}}$ such rescalings, and lower levels of the recursion will potentially be affected by rescalings from all higher levels. To ensure that all maintained vectors have $\polylog(n)$ bit complexity, we introduce tiny auxiliary coordinates that cancel between the residual and outgoing vector. Their magnitudes track subsequent rescaling, allowing the residual-energy bound to control error amplification.
These coordinates also provide the lower bounds on recursive norm scales needed to bound the number of corrections.
Together, these guarantees give $\frac{1}{\eta^2}\cdot\polylog(n)$ corrections per learner and polylogarithmically many bits per stored scalar.

Taking $\eta=\O{\frac{\eps}{H}}$ leaves room for errors from recursion, verification, and rounding.
By induction over the hierarchy, we conclude that the algorithm returns a $(1+\eps)$-approximation for all estimates simultaneously throughout the stream with probability at least $1-\delta$.
Since only $\O{H}=\O{\log n}$ sketches are active at any given time $t \in [m]$, the total space is $\frac{1}{\eps^2}\cdot\polylog(n)$ bits, as desired. 

\section{Preliminaries}
\begin{definition}[Differential privacy]
\label{def:dp}
Given a privacy parameter $\eps\ge0$ and $\delta\in[0,1)$, we say a randomized mechanism $M$ is $(\eps,\delta)$-differentially private if, for every ordered pair of neighboring datasets $X,X'$ differing in one sample and every output event $A$,
\begin{equation}
\label{eq:dp-definition}
\PPr{M(X)\in A}\le e^\eps\cdot\PPr{M(X')\in A}+\delta.
\end{equation}
\end{definition}
If $\delta=0$, the mechanism satisfies pure privacy, otherwise it satisfies approximate privacy. 
For an interactive mechanism, the output is its completed transcript, and the inequality must hold for every fixed analyst rule. 

\begin{definition}[Laplace distribution]
For $b>0$, the Laplace distribution $\Lap(b)$ has probability density $p(z)=\frac{1}{2b}\exp\left(-\frac{|z|}{b}\right)$. 
In particular, if $Z\sim\Lap(b)$, then for all $s\ge 0$, we have $\PPr{|Z|>s}=\exp\left(-\frac{s}{b}\right)$. 
\end{definition}

\begin{definition}[Global sensitivity]
A function $h:\calU^r\to\mathbb R$ has \emph{global sensitivity} at most $\Delta$ if $|h(S)-h(S')|\le\Delta$ for every pair of datasets $S,S'$ differing in one sample.
\end{definition}

\begin{lemma}[Laplace mechanism
{\cite[Theorem~3.6]{DworkR14}}]
\label{lem:laplace}
If $h$ has global sensitivity at most $\Delta$, then the
mechanism that releases $h(S)+Z$ for $Z\sim\Lap(b)$ with $Z$ drawn independently, is $(\Delta/b,0)$-differentially private.
\end{lemma}

\begin{lemma}[Post-processing and independent mixtures
{\cite[Definition~3 and Proposition~1]{DesfontainesP20}}]
\label{lem:dp-closure}
The following properties hold.
\begin{enumerate}
\item
If $M$ is $(\eps,\delta)$-differentially private, then any
computation using only $M(S)$ and independent randomness
is also $(\eps,\delta)$-differentially private.
In particular, clipping and rounding preserve privacy.

\item
Suppose $M_a$ is $(\eps,\delta)$-differentially private for
every fixed $a$.
Sampling $A$ independently of the dataset and then running
$M_A(S)$ with fresh randomness also gives an
$(\eps,\delta)$-differentially private mechanism.
\end{enumerate}
\end{lemma}

The second property allows us to prove a uniform privacy
guarantee after fixing auxiliary randomness independent of
the protected samples, and then average over that randomness.

\paragraph{\textsc{AboveThreshold}.}
Fix a threshold $\tau$ and a noise scale $b>0$.
Draw $Z\sim\Lap(2b)$ once.
For each test function $h_t:\calU^r\to\mathbb R$, draw fresh noise $V_t\sim\Lap(4b)$ and release $1$ if $h_t(S)+V_t\ge\tau+Z$, and $0$ otherwise. 
The procedure stops at the first $1$. 
The analyst may choose each test function and may stop earlier using previous outputs and independent randomness. 
All noise draws are independent; the same $Z$ is reused throughout the procedure.

\begin{lemma}[Privacy of \textsc{AboveThreshold}
{\cite[Theorem~3.23]{DworkR14}}]
\label{lem:above-threshold}
If every test function has global sensitivity at most $\Delta$ for every fixed preceding history, then the completed transcript of \textsc{AboveThreshold} is $(\Delta/b,0)$-differentially private, regardless of the number of tests.
\end{lemma}

\paragraph{Composition.}
We use the following form of advanced composition, which
also permits interleaved interactions: one mechanism may
remain active while another produces outputs.

\begin{lemma}[Advanced and concurrent composition \cite{BunS16,VadhanW21}]
\label{lem:dp-composition}
Suppose at most $m$ mechanisms access the same dataset, use independent randomness, and each is $(\eps_0,0)$-differentially private against every adaptive analyst.
Their interactions may be interleaved. 
The choice of mechanism, its inputs, and when to stop may depend on previous outputs and independent randomness, with no other access to the dataset. 
For every $\delta\in(0,1)$, the combined transcript is $\left(\eps_0\sqrt{2m\ln(1/\delta)} +\frac{m\eps_0^2}{2},\ \delta\right)$-differentially private. 
\end{lemma}

\section{Adaptive Data Analysis}
\label{sec:ada:gen}
Given $p$ independent blocks of samples from an unknown distribution $D$, for each $j\in[p]$, let $S_j=(X_{j,1},\ldots,X_{j,n_j})$ be the samples for the $j$-th block. 
We give each sample point a weight $w_{j,i}$, which corresponds to the scaling of $f_t(X_{j,i})$ for an output estimate by our mechanism, as illustrated in \Cref{alg:output}. 

\begin{algorithm}[H]
\caption{ADA mechanism for $k$ adaptive queries}
\label{alg:output}
\begin{algorithmic}[1]
\Require{sample blocks $S_1,\ldots,S_p$, query budget $k$, disagreement threshold $\tau$, Laplace noise scale $b$}
\State{Initialize $w_{j,i}\gets1/n_j$ for all $i\in[n_j]$, $j\in[p]$}
\For{$t=1,\ldots,k$}
\State{Receive query $f_t:\mathcal U\to[0,1]$}
\State{\Call{UpdateWeights}{$f_t$}}
\Comment{\Cref{alg:update} updates the weights $\{w_{j,i}\}$}
\State{Draw $E_t\sim\Lap(b)$ and $a_t\gets\sum_{i=1}^{n_1} w_{1,i}f_t(X_{1,i})+E_t$}
\State{Clip $a_t$ to $[0,1]$, round to the grid of spacing $\tau/64$}
\State{\Return $a_t$}
\EndFor
\end{algorithmic}
\end{algorithm}

The weights of each block sum to one, i.e., $\sum_{i\in[n_j]}w_{j,i}=1$ for all $j\in[p]$. 
We specify the last block $S_p$ to be the \emph{anchor}, whose weights shall remain uniform throughout the course of the queries. 
We shall set a threshold $\tau$ for the allowable discrepancy between outputs by blocks in the hierarchy, as well as a noise scale $b=\O{\frac{\tau}{\log n}}$. 
When the discrepancy exceeds $\tau$, we use \Cref{alg:update} to update the weights. 

\paragraph{Weight computation in \Cref{alg:update}.}
For each $j\in[p]$, let $f(S_j)=(f(X_{j,i}))_{i=1}^{n_j}$. 
We use the update rule 
\begin{equation}
\label{eq:weights}
w_j\gets\argmax_{\substack{w\ge0,\;\sum_iw_i=1\\w_i\le8/n_j}}\left(\frac{\tau}{4}\langle G_j,w\rangle-\frac{n_j}{2}\left\|w-\frac{\mathbf1}{n_j}\right\|_2^2\right).
\end{equation}
Here, $G_j$ is a score vector that informally keeps track of which sample points should receive more or less weight, to bring the answer of block $j$ into better agreement with those of its neighboring blocks. 
We also have a penalty term to enforce that the weights should remain near-uniform. 

\begin{algorithm}[H]
\caption{\textsc{UpdateWeights}$(f)$}
\label{alg:update}
\begin{algorithmic}[1]
\Require{$G_j$ (initialized to $\mathbf{0}$ before first query) and $Z_j\sim\Lap(2b)$ for $j\in[p]$}
\State{$j\gets1$}
\While{$j < p$}
\State{$q\gets\Call{Compare}{j,f}$}
\Comment{$q\in\{-1,0,+1\}$; defined below}
\State{\textbf{if} $q=0$ \textbf{then} $j\gets j+1$; \textbf{continue}}
\State{$G_j\gets G_j+q\,f(S_j)$}
\State{\textbf{if} $j+1< p$ \textbf{then} $G_{j+1}\gets G_{j+1}-\tfrac14q\,f(S_{j+1})$}
\State{Update all weights $\{w_{j,i}\}$ using~\Cref{eq:weights}}
\State{Draw $Z_j\sim\Lap(2b)$; $j\gets1$}
\Comment{Refresh this edge, then restart the scan}
\EndWhile
\end{algorithmic}
\end{algorithm}

\paragraph{Private comparison.}
We also use a subroutine $\textsc{Compare}(j,f)$ to privately test whether the discrepancy between two adjacent blocks exceeds the threshold $\tau$. 
The approach is essentially the standard \textsc{AboveThreshold} procedure from differential privacy, which adds Laplacian noise to both the difference between the output of the adjacent blocks and the threshold. 
Formally, it tests $q=+1,-1$ in that order, drawing fresh $V\sim\Lap(4b)$ for each test and returning the first sign that satisfies
\[q\left(\sum_{i=1}^{n_{j+1}}w_{j+1,i}f(X_{j+1,i})-\sum_{i=1}^{n_j}w_{j,i}f(X_{j,i})\right)+V\ge\tau+Z_j,\]
or $0$ if both tests fail.

\subsection{Reference Weights}
\label{sec:ada:ref:weights}
We first show the existence of accurate reference weights, which are near-uniform weights that enable the sample to simultaneously answer all $k$ queries correctly. 
Although we will not compute these weights explicitly, we will use the existence of these reference weights to show convergence and correctness of our adaptive data analysis mechanism. 

\begin{lemma}[Existence of accurate reference weights]
\label{lem:comparator}
Let $\eps\in[0,1]$ and $\alpha,\xi,\delta\in(0,1)$. 
Let $S=(X_1,\ldots,X_r)\sim D^r$ on a finite domain. 
Fix an adaptive analyst, and let $\mathsf{T}=M(S)$ be the (finite-valued) completed transcript of a $(\eps,\delta)$-private mechanism interacting with this analyst. 
Let $f_1(\mathsf{T},\cdot),\ldots,f_k(\mathsf{T},\cdot):\calU\to[0,1]$. 
For $\delta\le\frac{\alpha\xi}{32}$, there exists sample complexity $r=\O{\frac{1}{\alpha^2}\log\frac{k}{\xi}}$ such that with probability at least $1-\xi$, there exists a single vector $u\in\mathbb{R}^r$ such that
\begin{align*}
u_i\ge0,\qquad \sum_i u_i=1,\qquad u_i\le\frac8r,\qquad R(u):=\frac r2\sum_i(u_i-1/r)^2\le\eps^2,
\end{align*}
and for every $t\le k$,
\begin{equation}\label{eq:comparator-accuracy}
\left|\sum_i u_i f_t(\mathsf{T},X_i)-Df_t(\mathsf{T},\cdot)\right|\le\alpha.
\end{equation}
\end{lemma}
\begin{proof}
It suffices to take $r$ samples, for $r\ge\frac{1024}{\alpha^2}\ln\left(\frac{4(k+1)}{\xi}\right)$. 
We first draw $S\sim D^r$ and run the mechanism, obtaining the joint distribution of $(S,\mathsf{T})$. 
We then analyze this distribution by revealing $\mathsf{T}$ followed by $X_1,\ldots,X_r$. 
This order of analysis does not change how the samples or transcript are generated.

\paragraph{Controlling the bias from conditioning.}
Let $X_{<i}=(X_1,\ldots,X_{i-1})$ and $Y_i=(\mathsf{T},X_{<i})$.
For each possible value $y$ of $Y_i$, define
\[m_i(y)=\PPr{Y_i=y},\qquad p_i(x\mid y)=\frac{\PPr{X_i=x,Y_i=y}}{m_i(y)}\]
whenever $m_i(y)>0$; define $p_i(\cdot\mid y)$ arbitrarily otherwise. 
For ease of notation, we abbreviate $p_i(x):=p_i(x\mid Y_i)$, so that $p_i$ is the conditional distribution of the next sample, given the transcript and the samples already revealed. 
Note that unlike $D$, the distribution of $p_i$ may depend on both the transcript and samples.

We use differential privacy to upper bound how far $p_i$ can deviate from $D$. 
Our goal is to construct weights that undo the bias introduced by conditioning on the transcript. 
Ideally, we would assign each value $x$ the correction factor $D(x)/p_i(x)$, since multiplying its conditional probability $p_i(x)$ by this factor recovers $D(x)$ whenever $p_i(x)>0$. 
Unfortunately, this ratio may be arbitrarily large or small. 
To obtain reference weights close to uniform, we will clip these correction factors to the interval $[e^{-\eps},e^\eps]$ before normalizing. 
However, clipping leaves some bias uncorrected when $p_i(x)>e^\eps D(x)$ or $D(x)>e^\eps p_i(x)$.  
Thus, we now show that approximate differential privacy upper bounds the total excess in each of these inequalities by $\delta$ on average. 
To do this, we compare the original samples with one in which the $i$-th sample is independently replaced with a random sample.  

Draw an independent sample $X_i'\sim D$, and let $S^{(i)}$ be obtained by replacing $X_i$ with $X_i'$. 
Run the mechanism on $S^{(i)}$ with fresh randomness and the same analyst rule, producing a transcript $\mathsf{T}'$. 
The queries in $\mathsf{T}'$ need not be the same as those in $\mathsf{T}$. 
Let $P_i$ denote the distribution of $(\mathsf{T},X_{<i},X_i)$ and $Q_i$ denote the distribution of $(\mathsf{T}',X_{<i},X_i)$. 

For fixed $S=s$ and $X_i'=x'$, define $s'$ to be the dataset $s$ where the $i$-th sample is replaced by $x'$, so that the datasets $s$ and $s'$ are neighboring. 
Thus for any event $A$, differential privacy gives
\[\PPr{(M(s),s_{<i},s_i)\in A}\le e^{\eps}\cdot\PPr{(M(s'),s_{<i},s_i)\in A}+\delta,\]
and the same inequality in the other direction. 
Thus, averaging over $(S,X_i')$ gives
\begin{equation}
\label{eq:joint-dp}
P_i(A)\le e^{\eps} Q_i(A)+\delta,
\qquad
Q_i(A)\le e^{\eps} P_i(A)+\delta.
\end{equation}
For each possible value $y=(t,x_{<i})$ and sample value $x$, let
\[P_i(y,x):=\PPr{Y_i=y,\ X_i=x},\qquad Q_i(y,x):=\PPr{(\mathsf{T}',X_{<i})=y,\ X_i=x}.\]
Thus $P_i(y,x)$ and $Q_i(y,x)$ denote the probabilities of individual outcomes under the two experiments. 
By the definition of conditional probability, $P_i(y,x)=m_i(y)p_i(x\mid y)$. 
Since $X'_i$ is an independent draw, then the pair $(\mathsf{T}',X_{<i})$ has the same distribution as $Y_i$, but is independent of the original $X_i\sim D$, we have $Q_i(y,x)=m_i(y)D(x)$. 

Write $(z)_+=\max(0,z)$. 
Applying the first inequality in \Cref{eq:joint-dp} to $A_+=\{(y,x):P_i(y,x)>e^{\eps} Q_i(y,x)\}$, we have
\begin{align}
\Ex{\sum_x(p_i(x)-e^\eps D(x))_+}&=\sum_{y,x}(P_i(y,x)-e^\eps Q_i(y,x))_+\notag\\
&=P_i(A_+)-e^\eps Q_i(A_+)\le\delta.
\label{eq:posterior-upper}
\end{align}
Similarly, applying the second inequality to $A_-=\{(y,x):Q_i(y,x)>e^\eps P_i(y,x)\}$, we have
\begin{equation}
\label{eq:posterior-lower}
\Ex{\sum_x(D(x)-e^\eps p_i(x))_+}\le\delta.
\end{equation}
We remark that although we require privacy for the output transcript of the overall mechanism, the resulting bounds on conditional bias hold in expectation over $Y_i$, rather than separately for every transcript and prefix. 

\paragraph{Constructing the reference weights.} 
Now, if we could multiply the conditional mass $p_i(x)$ by $D(x)/p_i(x)$, we would recover the original mass $D(x)$. 
To achieve near-uniform weights, we instead use a bounded version of this correction:
\[v_i(x)=\min\left(e^{\eps},\max\left(e^{-\eps},\frac{D(x)}{p_i(x)}\right)\right),\qquad v_i=v_i(X_i),\]
where $v_i(x)=e^{\eps}$ when $p_i(x)=0$. 
Importantly, the error remaining after this correction is
\begin{align*}
d_i &:=\sum_x|p_i(x)v_i(x)-D(x)|=e^{-\eps}\sum_x(p_i(x)-CD(x))_+ +\sum_x(D(x)-Cp_i(x))_+\le2\delta
\end{align*}
by \Cref{eq:posterior-upper,eq:posterior-lower}. 

Condition on $Y_i=(\mathsf{T},X_{<i})$. 
This fixes the query $f_t(\mathsf{T},\cdot)$ and the correction function $v_i(\cdot)$, while $X_i$ has conditional distribution $p_i$. 
Writing $Df_t=\sum_xD(x)f_t(\mathsf{T},x)$, we have
\begin{align*}
\left|\Ex{v_i f_t(\mathsf{T},X_i)\mid Y_i}-Df_t\right|&=\left|\sum_x(p_i(x)v_i(x)-D(x))f_t(\mathsf{T},x)\right|\le d_i,
\end{align*}
because $0\le f_t\le1$.
Thus $d_i$ upper bounds the conditional bias of the corrected contribution for each of the $k$ queries.

\paragraph{Concentration and normalization.}
Although the transcript fixed the queries, the sample values may not be fixed and thus it remains to upper bound the deviation of the weighted values from the conditional expectation. 
To that end, for each $t=0,\ldots,k$, define
\[Z_{i,t}=v_i f_t(\mathsf{T},X_i)-\Ex{v_i f_t(\mathsf{T},X_i)\mid Y_i},\]
so that conditioned on $Y_i=(\mathsf{T},X_{<i})$, the random variable $Z_{i,t}$ has mean zero and magnitude at most $e^\eps$, since $0\le v_i f_t(\mathsf{T},X_i)\le e^\eps$. 
By similar reasoning, we have 
\[\Ex{Z_{i,t}^2\mid Y_i}\le e^{2\eps},\qquad\sum_{i=1}^r\Ex{Z_{i,t}^2\mid Y_i}\le re^{2\eps}.\]
Thus by Freedman's inequality,
\begin{align*}
\PPr{\left|\frac1r\sum_{i=1}^r Z_{i,t}\right|\ge\frac{\alpha}{8}}&\le2\exp\left(-\frac{(r\alpha/8)^2}{2\left(re^{2\eps}+e^\eps r\alpha/24\right)}\right)\\
&\le 2\exp\left(-\frac{r\alpha^2}{1024}\right),
\end{align*}
By our choice of $r$ and a union bound, we have that simultaneously for all $t=0,1,\ldots,k$,
\[\PPr{\left|\frac1r\sum_{i=1}^r Z_{i,t}\right|\le\frac{\alpha}{8}}\ge1-\frac{\xi}{2}.\]
Hence, by combining with the previous conditional bias analysis, we have that with probability at least $1-\xi$, simultaneously for all $t=0,1,\ldots,k$,
\begin{align*}
\left|\frac1r\sum_{i=1}^r v_i f_t(\mathsf{T},X_i)-Df_t\right|&\le\left|\frac1r\sum_{i=1}^r Z_{i,t}\right|+\frac1r\sum_{i=1}^r\left|\Ex{v_i f_t(\mathsf{T},X_i)\mid Y_i}-Df_t\right|\\
&\le\left|\frac1r\sum_{i=1}^r Z_{i,t}\right|+\frac1r\sum_{i=1}^r d_i\\
&\le\frac{\alpha}{4}.
\end{align*}
It remains to normalize the weights.
For the constant query $f_0\equiv1$, the previous bound gives $|\bar v-1|\le\frac{\alpha}{4}$, for $\bar v=\frac1r\sum_{i=1}^r v_i$. 
Thus the initial weights $v_i/r$ already have sum close to one. 
Moreover, after normalizing $u_i=\frac{v_i}{r\bar v}$, we have by triangle inequality, 
\begin{align*}
\left|\sum_{i=1}^r u_i f_t(\mathsf{T},X_i)-Df_t\right|&\le\frac{\left|r^{-1}\sum_i v_i f_t(\mathsf{T},X_i)-Df_t\right|+|1-\bar v|Df_t}{\bar v}\\
&\le\frac{\alpha/2}{1-\alpha/4}\le\alpha.
\end{align*}
Thus, concentration holds for the normalized weights. 

\paragraph{Bounding the deviation from uniform weights.}
It remains to bound the deviation of $u$ from uniform weights, as measured by $R(u):=\frac r2\sum_{i=1}^r\left(u_i-\frac1r\right)^2$. 
We show that $R(u)\le\eps^2$. 
By construction of the clipping, we have $v_i,\bar v\in[e^{-\eps},e^{\eps}]$. 
Hence,
\[u_i\le\frac{e^{2\eps}}{r}\le\frac{e^2}{r}<\frac8r.\]
Also, $(v_i-e^{-\eps})(v_i-e^{\eps})\le0$ implies $v_i^2\le(e^{\eps}+e^{-\eps})v_i-1$. 
Thus by averaging across all $v_i$, we similarly have $\overline{v^2}\le(e^{\eps}+e^{-\eps})\bar v-1$. 
Since $u_i=v_i/(r\bar v)$ and $r^{-1}\sum_i v_i=\bar v$, then by completing the square,
\begin{align*}
2R(u)&=\frac1r\sum_i\left(\frac{v_i}{\bar v}-1\right)^2=\frac{\overline{v^2}}{\bar v^{\,2}}-1\\
&\le\frac{e^{\eps}+e^{-\eps}}{\bar v}
     -\frac1{\bar v^{\,2}}-1\\
&=\frac{(e^{\eps}-e^{-\eps})^2}{4}
  -\left(\frac1{\bar v}-\frac{e^{\eps}+e^{-\eps}}2\right)^2\\
&\le\frac{(e^{\eps}-e^{-\eps})^2}{4}.
\end{align*}
Finally, convexity of $e^x-e^{-x}$ on $[0,1]$ gives
\[e^\eps-e^{-\eps}\le(e-e^{-1})\eps\le2\sqrt2\,\eps.\]
Thus $R(u)\le\eps^2$, completing the proof.
\end{proof}

\subsection{Progress and Convergence}
\label{sec:ada:progress}
Although we do not compute the reference weights established in \Cref{sec:ada:ref:weights}, their existence will help us establish a bounded non-negative potential function so that each update to the weights maintained by the algorithm will make progress, by decreasing the potential. 
To avoid a circular argument where we need privacy to bound the number of corrections and a bound on the number of corrections to establish privacy, we first analyze a version with predetermined correction limits. 
These limits allow us to establish privacy first; the progress analysis then shows that they are not reached with high probability.

Let $\alpha,\delta<\frac{1}{10}$ be the accuracy and failure parameters. 
Choose $p$ to be the smallest integer with $4^{p-2}(\alpha/(4p))^2\ge1$, and set the threshold to be $\tau=\frac{\alpha}{4p}$, so that $p=\O{\log\frac{1}{\alpha}}$ and $1\le4^{p-2}\tau^2<4$. 
For every $j\in[p]$, including the anchor, let $\mathcal W_j$ and $R_j$ be the feasible set and quadratic penalty in \Cref{eq:weights}. 
Let $\mu_j(f)=\langle w_j,f(S_j)\rangle$ be the current prediction, including the uniform anchor. 
Define edge $j$ so that it joins blocks $j$ and $j+1$. 

For each block $j<p$, we allocate total privacy budget $\eps_j=\frac1{2^{j-1}\sqrt p}$ over the course of the $k$ adaptive interactions. 
For the anchor, we set the total privacy budget to be $\eps_p=\frac{\tau}{512}$. 
Since \Cref{lem:comparator} shows the reference weights satisfy $R_j(u_j)\le\eps_j^2$, the choice of $\eps_j$ provides upper bounds on the initial potential induced by $R_j$. 
The anchor's smaller privacy budget will ensure that its accurate reference weights are sufficiently close to uniform that the uniform weights themselves are accurate.

To establish privacy before proving an upper bound on the number of corrections, we analyze an auxiliary version with predetermined correction limits. 
Let $T_j$ count the corrections between blocks $j$ and $j+1$, and for $j\in[1,p)$, set $B_j=\left\lceil\frac{32}{4^{j-1}\tau^2}\right\rceil$ to be the correction limit for $j$. 
That is, whenever a correction makes $T_j=B_j$, the auxiliary version performs the correction and then stops accessing the data, thus allowing us to establish privacy. 
We will show that $T_j<B_j$ with high probability for all $j\in[1,p)$. 
For the analysis, let $\mathsf T$ record all internal comparison outcomes together with the rounded answers. 
This augmented transcript is finite-valued because the correction limits upper bound the number of comparisons and the answers lie on a finite grid. 

\begin{proposition}[Reference accuracy under bounded privacy]
\label{prop:good}
Suppose that, for every $j\in[p]$, the transcript $\mathsf T$ is $(\eps_j,\delta)$-private with respect to block $S_j$, and that $n_j$ and $\delta$ satisfy the assumptions of \Cref{lem:comparator} with accuracy $\tau/64$ and failure probability $\xi/(p+1)$.
Suppose also that the probability of any noise draw having magnitude greater than $\tau/64$ is at most $\xi/(p+1)$. 
Then, with probability at least $1-\xi$, the following properties hold simultaneously:
\begin{enumerate}
\item 
For each $j<p$, there exists a single vector $u_j\in\mathcal W_j$ such that
\[|\langle u_j,f_t(S_j)\rangle-Df_t|\le\frac{\tau}{64}\quad\text{for every }t\in[k],\qquad R_j(u_j)\le\eps_j^2.\]
\item 
The uniform anchor satisfies
\[|\mu_p(f_t)-Df_t|\le\frac{\tau}{32}\quad\text{for every }t\in[k].\]
\item 
Every noise draw during the execution has magnitude at most $\tau/64$.
\end{enumerate}
Moreover, these reference vectors satisfy
\begin{equation}\label{eq:initial-budget}
P:=\sum_{j=1}^{p-1}4^{j-1}R_j(u_j)\le\sum_{j=1}^{p-1}4^{j-1}\eps_j^2=\frac{p-1}{p}\le1.
\end{equation}
\end{proposition}
\begin{proof}
Apply \Cref{lem:comparator} to each block with accuracy $\tau/64$, privacy parameter $\eps_j$, and failure probability $\xi/(p+1)$. 
By hypothesis, the required sample-size bound holds and
\[\delta\le\frac{(\tau/64)(\xi/(p+1))}{32}=\frac{\tau\xi}{2048(p+1)}.\]
When applying the lemma to block $S_j$, we regard the other independent blocks as part of the mechanism's randomness. 
This gives the claimed reference weights for every $j<p$. 

For the anchor, the lemma gives a reference vector $u_p$ with error at most $\tau/64$ and $R_p(u_p)\le\eps_p^2$. 
The anchor uses uniform weights, so $\mu_p(f)=n_p^{-1}\sum_i f(X_{p,i})$. 
Since both weight vectors sum to one, $\sum_i(1/n_p-u_{p,i})=0$. 
Thus subtracting $1/2$ from each query value does not change the difference between
their predictions:
\begin{align*}
|\mu_p(f)-\langle u_p,f(S_p)\rangle|&=\left|\sum_i\left(\frac1{n_p}-u_{p,i}\right)\left(f(X_{p,i})-\frac12\right)\right|\\
&\le\frac12\|u_p-\mathbf1/n_p\|_1\le\frac{\sqrt{n_p}}2\|u_p-\mathbf1/n_p\|_2\\
&=\sqrt{R_p(u_p)/2}\le\eps_p=\frac{\tau}{512}.
\end{align*}
Here, the first inequality uses $|f(X_{p,i})-1/2|\le\frac{1}{2}$, and the second is Cauchy-Schwarz.
Thus by triangle inequality, the anchor's error is at most $\tau/64+\tau/512<\tau/32$ for every selected query. 

Each application of the lemma fails with probability at most $\xi/(p+1)$. 
By assumption, the same bound holds for the event that any noise draw exceeds $\tau/64$ in magnitude.
A union bound over the $p$ blocks and the noise event therefore gives all three properties simultaneously with probability at least $1-\xi$.
Finally, the reference weights satisfy
\[P=\sum_{j=1}^{p-1}4^{j-1}R_j(u_j)\le\sum_{j=1}^{p-1}4^{j-1}\eps_j^2=\frac{p-1}{p}\le1.\]
\end{proof}

Let $\mathcal E$ denote the event that all conclusions of \Cref{prop:good} hold, so that under the corresponding hypotheses, $\PPr{\mathcal E}\ge1-\xi$. 
Fix an execution of the auxiliary mechanism in $\mathcal E$, and choose one reference vector $u_j$ for each $j<p$ satisfying \Cref{prop:good}. 
We use these same reference vectors throughout the execution. 
The argument below is deterministic.

\paragraph{One block.}
For a block of size $r$, let $\calW$ be its feasible set, $R$ be its penalty term, and  $w(G)$ be the weights chosen from scores $G$. 
Define the maximization objective
\begin{equation}
\label{eq:value}
\Phi(G)=\max_{w\in\mathcal W}\left(\frac{\tau}{4}\langle G,w\rangle-R(w)\right).
\end{equation}
Fix a reference vector $u\in\mathcal W$ throughout the updates. 
For a correction $g$, compare $u$ with the current weights $w(G)$ before updating them. 
Holding both vectors fixed, their scores increase by $\langle u,g\rangle$ and $\langle w(G),g\rangle$, respectively, while their penalties remain unchanged. 
We call this difference $\langle u-w(G),g\rangle$ the reference vector's score advantage.
We now upper bound the sum of these advantages as the algorithm updates its weights, with $u$ remaining fixed. 
We remark that in the following statement, the index $s$ counts the number of updates to the weights, rather than the number of queries. 

\begin{lemma}[One update and accumulated reference advantage]
\label{lem:regret}
For all $G,g\in\mathbb R^r$,
\begin{equation}\label{eq:smooth-value}
\Phi(G+g)-\Phi(G)\le\frac{\tau}{4}\langle w(G),g\rangle+\frac{\tau^2}{32}\|g\|_\infty^2,
\end{equation}
where $\|g\|_\infty=\max_i|g_i|$. 
For any sequence $G^{(0)}=0$, $G^{(s)}=G^{(s-1)}+g^{(s)}$, write $w^{(s)}=w(G^{(s-1)})$ for the weights before update $s$. 
Then every fixed $u\in\mathcal W$ satisfies
\begin{equation}\label{eq:regret}
\sum_{s=1}^m\langle u-w^{(s)},g^{(s)}\rangle\le\frac{4R(u)}{\tau}+\frac{\tau}{8}\sum_{s=1}^m\|g^{(s)}\|_\infty^2.
\end{equation}
\end{lemma}
\begin{proof}
Let $w=w(G)$ and $w'=w(G+g)$. 
The feasible segment from $w$ to $w'$ and optimality of $w$ give
\[\left\langle\frac{\tau}{4}G-\nabla R(w),w'-w\right\rangle\le0.\]
Expanding the quadratic penalty and completing the square, we have
\begin{align*}
\Phi(G+g)-\Phi(G)-\frac{\tau}{4}\langle w,g\rangle&\le\frac{\tau}{4}\langle w'-w,g\rangle-\frac{r}{2}\|w'-w\|_2^2=\frac{\tau^2}{32r}\|g\|_2^2-\frac r2\left\|w' - w-\frac{\tau}{4r}g\right\|_2^2\\
&\le\frac{\tau^2}{32r}\|g\|_2^2\le\frac{\tau^2}{32}\|g\|_\infty^2,
\end{align*}
which gives the first claim. 

Now, to obtain \Cref{eq:regret}, sum \Cref{eq:smooth-value} over the updates. 
The left-hand side telescopes, and $\Phi(0)=0$ because the uniform weights are feasible and have zero penalty. 
Hence, we have 
\[\Phi(G^{(m)})\le\frac{\tau}{4}\sum_{s=1}^m\langle w^{(s)},g^{(s)}\rangle+\frac{\tau^2}{32}\sum_{s=1}^m\|g^{(s)}\|_\infty^2.\]
Finally, by the definition of $\Phi$ and the feasibility of $u$, we have
\begin{align*}
\frac{\tau}{4}\left\langle\sum_{s=1}^m g^{(s)},u\right\rangle-R(u)&\le\Phi(G^{(m)}).
\end{align*}
Putting these inequalities together, rearranging, and multiplying by $\frac{4}{\tau}$ gives the desired claim of
\[\sum_{s=1}^m\langle u-w^{(s)},g^{(s)}\rangle\le\frac{4R(u)}{\tau}+\frac{\tau}{8}\sum_{s=1}^m\|g^{(s)}\|_\infty^2.\]
\end{proof}

\paragraph{Combining the blocks.}
We now define a potential $\Psi$ to track progress across all learned blocks. 
Let $\Phi_j$ denote \Cref{eq:value} for block $j$. 
The weights maintained by the algorithm have objective value $\Phi_j(G_j)$, while the reference weights $u_j$ have objective value $\frac{\tau}{4}\langle G_j,u_j\rangle-R_j(u_j)$. 
Define $\Psi$ to be the weighted sum of these objective differences:
\begin{align*}
\Psi:=\sum_{j=1}^{p-1}4^{j-1}\left[\Phi_j(G_j)-\left(\frac{\tau}{4}\langle G_j,u_j\rangle-R_j(u_j)\right)\right].
\end{align*}
Observe that each difference is non-negative because $\Phi_j(G_j)$ is the maximum over all feasible weights, including $u_j$ and thus $\Psi\ge0$. 
We remark that the multipliers $4^{j-1}$ compensate for the factor $\frac{1}{4}$ in the update rule for the scores of the neighboring block, so that the two updated blocks have  equal coefficients in the linear part of the progress calculation.

Initially $G_j=0$, so the uniform weights have objective value zero, the reference weights have value $-R_j(u_j)$. 
Thus the difference in their objective values is initially $R_j(u_j)$, giving
\begin{equation}\label{eq:initial-potential}
\Psi_{\mathrm{initial}}=\sum_{j=1}^{p-1}4^{j-1}R_j(u_j)\le1.
\end{equation}
The next proposition shows that every correction decreases $\Psi$ by at least some fixed amount. 
Since $\Psi$ starts at most one and remains non-negative, this will upper bound the number of corrections.

\begin{proposition}[Progress and correction bounds]
\label{lem:progress}\label{prop:corrections}
Conditioned on the event $\mathcal E$, every correction on edge $j$ in the auxiliary execution decreases $\Psi$ by at least $4^{j-1}\tau^2/16$. 
Consequently, every prefix satisfies
\begin{equation}\label{eq:weighted-corrections}
\sum_{j=1}^{p-1}4^{j-1}T_j\le\frac{16}{\tau^2}.
\end{equation}
In particular, $T_j\le16/(4^{j-1}\tau^2)$, the total number of corrections is at most $16/\tau^2$, and no correction limit $B_j$ is reached.
Moreover, block $1$ receives at most $16/\tau^2$ weight updates and each block $2\le j<p$ receives at most $80/(4^{j-1}\tau^2)$.
\end{proposition}
\begin{proof}
Consider a correction on query $f$ with sign $q$ at edge $j$, using the weights before the correction. 
A positive test $q=1$ and the assumption on the magnitude of the noise imply
\begin{equation}\label{eq:certified-gap}
q\left(\mu_{j+1}(f)-\mu_j(f)\right)\ge\tau-\frac{2\tau}{64}\ge\frac{3\tau}{4}.
\end{equation}

First, suppose $j<p-1$. 
The outputs of the two reference weights differ by at most $\tau/32$, since each is within $\tau/64$ of $Df$. 
Hence their combined advantage over the current weights satisfies
\[q\left(\langle u_j-w_j,f(S_j)\rangle-\langle u_{j+1}-w_{j+1},f(S_{j+1})\rangle\right)\ge\frac{3\tau}{4}-\frac{\tau}{32}\ge\frac{\tau}{2}.\] 
We change the scores by the increments $qf(S_j)$ and $-qf(S_{j+1})/4$, respectively. 
By \Cref{eq:smooth-value}, the change in $\Psi$ is at most
\[-\frac{4^{j-1}\tau^2}{8}+\frac{\tau^2}{32}\left(4^{j-1}+\frac{4^j}{16}\right)\le-\frac{4^{j-1}\tau^2}{16}.\]

For the last edge $j=p-1$, only block $p-1$ is updated. 
By \Cref{prop:good}, its reference prediction differs from the uniform anchor by at most $\tau/64+\tau/32$, so
\[q\langle u_{p-1}-w_{p-1},f(S_{p-1})\rangle\ge\frac{3\tau}{4}-\frac{\tau}{64}-\frac{\tau}{32}\ge\frac{\tau}{2}.\]
Thus, the change in $\Psi$ is at most
\[-\frac{4^{p-2}\tau}{4}\cdot\frac{\tau}{2}+\frac{4^{p-2}\tau^2}{32}=-\frac{3}{32}4^{p-2}\tau^2\le-\frac{4^{p-2}\tau^2}{16}.\]
Summing the decrease over all corrections performed so far and using $\Psi_{\mathrm{initial}}\le1$ and $\Psi\ge0$, we have
\[0\le\Psi\le1-\frac{\tau^2}{16}\sum_{j=1}^{p-1}4^{j-1}T_j,\]
proving \Cref{eq:weighted-corrections}. 
Each term gives $T_j\le16/(4^{j-1}\tau^2)$, and the total correction bound follows because every multiplier is at least one. 
If a first correction limit were reached, the prefix including that correction would contradict
\[T_j\le\frac{16}{4^{j-1}\tau^2}<B_j.\]
Thus no limit is reached.

Finally, block $1$ is updated only on edge $1$. 
For $2\le j<p$, block $j$ is updated only on its two incident edges, giving at most
\[T_{j-1}+T_j\le\frac{16}{4^{j-2}\tau^2}+\frac{16}{4^{j-1}\tau^2}=\frac{80}{4^{j-1}\tau^2}\]
weight updates.
\end{proof}

Run the auxiliary and original mechanisms with the same sample blocks and random choices. 
Their executions agree until a correction limit is reached. 
Since no limit is reached on $\mathcal E$, their executions agree on this event, and the correction bounds also hold for the original mechanism.
\begin{proposition}[Accuracy on the good event]
\label{prop:accuracy}
Conditioned on the event $\mathcal E$, the original mechanism completes all $k$ queries, and every public answer has population error less than $\alpha$.
\end{proposition}
\begin{proof}
The correction bound in \Cref{prop:corrections} ensures that every query finishes. 
During the final scan, the weights remain unchanged and both signed tests fail at every edge. 
The assumption on the magnitude of the noise thus gives
\[|\mu_{j+1}(f)-\mu_j(f)|\le\tau+\frac{2\tau}{64}=\frac{33\tau}{32}.\]
Clipping cannot increase error relative to $Df\in[0,1]$. 
Combining the adjacent discrepancies, the anchor bound from
\Cref{prop:good}, and the errors from output noise and rounding, we obtain
\begin{align*}
|a-Df|&\le\sum_{j=1}^{p-1}|\mu_j(f)-\mu_{j+1}(f)|+\frac{\tau}{32}+\frac{\tau}{64}+\frac{\tau}{128}\\
&\le(p-1)\frac{33\tau}{32}+\frac{7\tau}{128}<4p\tau=\alpha.
\end{align*}
\end{proof}

\subsection{Privacy}
\label{sec:privacy}
We now verify the hypotheses of \Cref{prop:good} using only the predetermined correction limits. 
This argument applies to every execution, including those that reach a limit; it does not assume the event $\mathcal E$ or the progress bound.

\paragraph{Sensitivity of the weighted predictions.}
The global sensitivity is the largest change in a function's value when one sample is replaced. 
We upper bound the global sensitivity, accounting for changes in both the query values and the weights.

\begin{lemma}[Projection and sensitivity]
\label{lem:sensitivity}
For a block of size $r$, the weights in \Cref{eq:weights} satisfy
\begin{equation}\label{eq:projection}
w_i(G)=\frac1r\operatorname{clip}_{[0,8]}
\left(\frac{\tau}{4}G_i+c\right),
\end{equation}
where $\operatorname{clip}_{[0,8]}(x)=\min(8,\max(0,x))$ and $c$ is chosen so that the weights sum to one. 
For fixed functions $G:\calU\to\mathbb R$ and $f:\calU\to[0,1]$, the weighted prediction computed from scores $G(X_1),\ldots,G(X_r)$ has global sensitivity at most $16/r$.
\end{lemma}
\begin{proof}
The sum of the weights in \Cref{eq:projection} is continuous and non-decreasing in $c$, with limits zero and eight, so there exists $c$ for which the weights sum to one. 
To prove that these weights solve \Cref{eq:weights}, let $v$ be any vector satisfying the constraints of that maximization, i.e., $0\le v_i\le\frac{8}{r}$ and $\sum_i v_i=1$. 
We will show that the objective at $w$ is at least its value at $v$.
To verify that these weights maximize the objective
\[\frac{\tau}{4}\langle G,w\rangle-R(w)=\frac{\tau}{4}\langle G,w\rangle-\frac{r}{2}\sum_i(w_i-1/r)^2,\]
observe that its partial derivative with respect to $w_i$ is $(\tau/4)G_i-rw_i+1$.
For the weights in \Cref{eq:projection}, this derivative equals $1-c$ when $0<w_i<8/r$, is at most $1-c$ when $w_i=0$, and is at least $1-c$ when $w_i=8/r$.
A feasible $v$ satisfies $v_i-w_i\ge0$ when $w_i=0$ and $v_i-w_i\le0$ when $w_i=8/r$.
Thus
\[\left\langle\frac{\tau}{4}G-r(w-\mathbf1/r),v-w\right\rangle\le(1-c)\sum_i(v_i-w_i)=0,\]
where the equality uses that both $v$ and $w$ sum to one. 
Expanding $R(v)=\frac{r}{2}\|(w-\mathbf1/r)+(v-w)\|_2^2$ gives
\begin{align*}
\left(\frac{\tau}{4}\langle G,v\rangle-R(v)\right)-\left(\frac{\tau}{4}\langle G,w\rangle-R(w)\right)&=\left\langle\frac{\tau}{4}G-r(w-\mathbf1/r),v-w\right\rangle-\frac{r}{2}\|v-w\|_2^2\\
&\le-\frac r2\|v-w\|_2^2\le0.
\end{align*}
Hence every feasible $v$ has objective value at most that of $w$. 
Thus the weights given by \Cref{eq:projection} solve the maximization in \Cref{eq:weights}.

Now replace sample $i$. 
Let $w,w'$ denote the original and recomputed weights, with respective normalizing shifts $c,c'$. 
For every $\ell\ne i$, the score $G_\ell$ is unchanged, so the formula for $w'_\ell$ differs from that for $w_\ell$ only by replacing $c$ with $c'$. 
Since clipping is non-decreasing, the weights with $\ell\ne i$ all move in the same direction. 
Their total gain or loss must be balanced by the change in weight at coordinate $i$, because both weight vectors sum to one. 
Thus
\[\sum_{\ell\ne i}|w_\ell-w_\ell'|=\left|\sum_{\ell\ne i}(w_\ell-w_\ell')\right|=|w_i-w_i'|\le\frac{8}{r}.\]
Since query values lie in $[0,1]$, the combined contribution of these unchanged samples changes by at most $8/r$. 
Finally, at coordinate $i$, both the old and new weighted query values lie in $[0,8/r]$, so their difference has magnitude at most $8/r$.
Adding these two bounds, it follows that the global sensitivity is at most $16/r$.
\end{proof}

We remark that for any fixed history, the current query and each score function $G_j$ are fixed. 
Thus \Cref{lem:sensitivity} gives sensitivity at most $16/n_j$ for each block $j$ and the same bound holds for the uniform anchor, with sensitivity at most $1/n_p$.

\paragraph{Privacy of releases and threshold phases.} 
It is well-known that the Laplace mechanism, i.e., adding $\Lap(\Delta/\eps)$ noise to a function with global sensitivity $\Delta$ is $\eps$-private, c.f., \Cref{lem:laplace}. 
Clipping and rounding preserve this guarantee by post-processing of differential privacy, c.f., \Cref{lem:dp-closure}. 

A \emph{phase} on edge $j$ begins when its threshold noise $Z_j\sim\Lap(2b)$ is drawn. 
During the phase, every signed test for $q$ on that edge reuses $Z_j$ and draws fresh noise $V\sim\Lap(4b)$; these tests may span multiple queries. 
The phase ends at the first test that triggers a correction or when the execution ends. 
Refreshing $Z_j$ after a correction starts a new phase.
Each phase is therefore an instance of \textsc{AboveThreshold}.
Holding the other blocks fixed, a phase on an edge incident to block $S_j$ is therefore $16/(n_jb)$-private with respect to that block~\cite[Theorem~3.23]{DworkR14}. 
This guarantee allows adaptively chosen test functions and termination without a successful test.

For privacy composition, we count each noisy answer as one application of the Laplace mechanism and each phase as one application of \textsc{AboveThreshold}. 
Each edge starts one phase initially and another after each correction, except that reaching a correction limit stops the execution before the threshold is refreshed. 
Thus edge $j$ uses at most $B_j$ phases. 
Counting the phases on edges incident to each block and, for block $1$, the at most $k$ noisy query answers, the number of private components accessing each block is at most
\begin{equation}\label{eq:component-counts}
C_1:=k+B_1,\qquad
C_j:=B_{j-1}+B_j\quad(2\le j<p),\qquad
C_p:=B_{p-1}.
\end{equation}
Choose failure probability $\delta=\frac{\tau\xi}{4096(p+1)}$, a logarithmic term $L=\ln\left[\frac{2p}{\delta}\left(k+\sum_{j=1}^{p-1}B_j+1\right)\right]$ to account for advanced composition, and noise parameter $b=\frac{\tau}{256L}$. 
Choose integer block sizes
\begin{align*}
n_1=\O{\left(\frac{\sqrt{k}}{\alpha}+\frac{1}{\alpha^2}\right)\log^4\frac{2k}{\alpha\xi}},\qquad n_j=\O{\frac{1}{\alpha^2}\log^4\frac{2k}{\alpha\xi}}\qquad(2\le j\le p),
\end{align*}
with sufficiently large constants so that $n_j\ge\left\lceil\max\left(\frac{128\sqrt{C_jL}}{b\eps_j},\,\frac{2^{22}L}{\tau^2}\right)\right\rceil$.
The first term ensures privacy; the second supplies enough samples for \Cref{lem:comparator}.

\begin{theorem}[Privacy and the hypotheses for progress]
\label{thm:privacy}
With these parameter choices, the auxiliary transcript is $(\eps_j,\delta)$-private with respect to replacing one sample in $S_j$, for every $j\in[p]$. 
Then the hypotheses of \Cref{prop:good} hold, so $\PPr{\mathcal E}\ge1-\xi$.
\end{theorem}
\begin{proof}
Fix a block $S_j$ and assign independent randomness to all potential phases and numerical releases in advance. 
Fix the other sample blocks, the analyst's independent randomness, and the randomness of components that do not directly access $S_j$. 
All this information is independent of $S_j$. 
Observe that any fixed history determines every other block's weighted query estimate. 
Thus the choice of component and its input function uses only previous outputs and this fixed information.

Now, there are at most $C_j$ components accessing $S_j$, each $16/(n_jb)$-private. 
A phase may remain active while other components produce outputs, so we use concurrent composition~\cite[Corollary~1.11]{VadhanW21}, which permits the usual composition bounds for pure differential privacy even when the interactions are interleaved. 
The sharpened advanced composition bound~\cite{BunS16} gives $(\eps_j,\delta)$-privacy because
\begin{equation}\label{eq:composition}
\frac{16}{n_jb}\sqrt{2C_j\ln(1/\delta)}+\frac{C_j}{2}\left(\frac{16}{n_jb}\right)^2\le\frac{\sqrt2}{8}\eps_j+\frac{\eps_j^2}{128L}<\eps_j.
\end{equation}
The inequalities follow from the choice of $n_j\ge\left\lceil\max\left(\frac{128\sqrt{C_jL}}{b\eps_j},\,\frac{2^{22}L}{\tau^2}\right)\right\rceil$, $L\ge\max\left(1,\ln\frac{1}{\delta}\right)$, and $\eps_j\le1$. 
Averaging over the fixed information preserves this guarantee by closure under data-independent mixtures, c.f., \Cref{lem:dp-closure}.

It remains to verify the sample-size and noise assumptions of \Cref{prop:good}.
Our choice of $L$ satisfies $L\ge\ln\frac{4(k+1)(p+1)}{\xi}$, so
\[n_j\ge\frac{1024}{(\tau/64)^2}\ln\frac{4(k+1)(p+1)}{\xi}, \qquad\delta\le\frac{(\tau/64)(\xi/(p+1))}{32}.\]
Thus \Cref{lem:comparator} applies with accuracy $\tau/64$ and failure probability $\xi/(p+1)$.

Each time the mechanism begins checking from edge $1$, it is either starting a query or restarting after a correction. 
There are at most $k$ queries and $\sum_j B_j$ corrections, so this happens at most $k+\sum_j B_j$ times. 
Before the next correction or the end of the query, it checks at most $p-1$ edges, with at most two signed tests per edge.
Thus the comparisons use at most $2(p-1)\left(k+\sum_j B_j\right)$ draws of test noise. 
There are also at most $p-1+\sum_j B_j$ threshold draws and at most $k$ draws for the noise added to query answers.
Thus the total number of noise draws is at most $2p\left(k+\sum_{j=1}^{p-1}B_j+1\right)$. 
Every fresh draw has scale at most $4b$, so its probability of exceeding $\tau/64$ in magnitude is at most $\exp\left(-\frac{\tau}{256b}\right)=e^{-L}$. 
By the setting of $L$ and a union bound, the conditional probability that any noise draw has magnitude greater than $\tau/64$ is at most $\delta\le\frac{\xi}{p+1}$. 
Hence all hypotheses of \Cref{prop:good} hold.
\end{proof}

Combining this theorem with \Cref{prop:corrections,prop:accuracy} gives the correction and accuracy guarantees for the original mechanism with probability at least $1-\xi$.

\subsection{Putting privacy, progress, and accuracy together}
\label{sec:conclusion}

We combine privacy of the auxiliary mechanism with the progress
and accuracy bounds to obtain the following guarantee for the
original mechanism.

\begin{theorem}[Accurate answers to adaptive queries]
\label{thm:ada:main}
For any accuracy parameter $\alpha>0$ and failure probability $\xi\in(0,1)$, there exists a randomized mechanism using $\O{\left(\frac{\sqrt{k}}{\alpha}+\frac{1}{\alpha^2}\right)\log^5\frac{k}{\alpha\xi}}$ independent samples from any unknown distribution $D$ on a finite domain $\calU$. 
Against any adaptive analyst submitting queries $f_t:\calU\to[0,1]$, with probability at least $1-\xi$, the mechanism returns all $k$ answers satisfying 
\[\left\lvert a_t-\EEx{X\sim D}{f_t(X)}\right\rvert\le\alpha.\]
\end{theorem}
\begin{proof}
Consider the mechanism in \Cref{alg:output,alg:update}, with the hierarchy and parameter choices specified above. 
By \Cref{thm:privacy}, the auxiliary mechanism satisfies $\PPr{\mathcal E}\ge1-\xi$. 
Run the auxiliary and original mechanisms with the same samples and random choices. 
Their executions agree until a correction limit is reached.
Conditioned on the event $\mathcal E$, no limit is reached by \Cref{prop:corrections}, so their executions agree throughout all $k$ queries. 
Thus \Cref{prop:corrections,prop:accuracy} give the claimed termination, accuracy, and correction bounds for the original mechanism.

For the sample complexity, summing the choices of $n_j$ in \Cref{thm:privacy} gives
\begin{align*}
N&=\O{\left(\frac{\sqrt{k}}{\alpha}+\frac{p}{\alpha^2}\right)\log^4\frac{2k}{\alpha\xi}}=\O{\left(\frac{\sqrt{k}}{\alpha}+\frac1{\alpha^2}\right)\log^5\frac{2k}{\alpha\xi}},
\end{align*}
where the second equality uses $p=\O{\log(2/\alpha)}$.
\end{proof}

We remark that up to logarithmic factors, the $\frac{\sqrt{k}}{\alpha}$ term comes from composing privacy over the $k$ noisy query answers, while the $\frac{1}{\alpha^2}$ term supplies enough samples for accurate reference weights and the weight corrections. 
In particular, when $k=\O{\frac{1}{\alpha^2}}$, the sample complexity is $\tO{\frac{1}{\alpha^2}}\cdot\polylog\left(\frac{1}{\xi}\right)$.

We also observe that for storage, each learned block maintains one score and one weight per sample. 
Recomputing its weights requires at most $\O{n_j}$ additional scalars. 
The remaining state consists of the threshold noise for each edge, the current indices, and the current answer. 
Neither reference vectors nor previous queries need to be stored. 
This gives the total bound of $\O{N+p}$ stored scalars, where $N$ is defined in the proof of \Cref{thm:ada:main}. 

As a corollary, we note that the randomized ADA mechanism can also be used to perform the following \textit{conditional release} task: for an interval request at time $t\in[m]$, the analyst provides a statistical query $f_t$ and an interval $[L_t,H_t]$, where $L_t\leq H_t$ may vary with $t$ and depend on previous responses. 
The mechanism privately tests whether the population mean of $f_t$ lies in this interval, up to additive error $\alpha$, and returns \Accept, \High, or \Low. 
Additionally, after a \High or \Low response, the analyst may request a value $a_t$ approximating $\EEx{X\sim D}{f_t(X)}$ to additive error $\alpha$, provided that the mechanism has not halted. 

\begin{corollary}[Adaptive conditional release]
\label{cor:conditional:release}
Fix an accuracy parameter $\alpha\in(0,1/10)$, a failure probability $\xi\in(0,1)$, and integer budgets $m\ge 1$ and $R,T\ge 0$. 
Suppose an adaptive analyst submits at most $m$ requests in total for queries $f_t:\calU\rightarrow[0,1]$, where each request is either an interval request $(f_t,[L_t,H_t])$ for adaptively chosen $L_t\leq H_t$ or a numerical request (at most $R$ such requests). 
There exists a randomized mechanism using $\O{\left(\frac{\sqrt{R+T}}{\alpha}+\frac{1}{\alpha^2}\right)\log^5\frac{m+R+T}{\alpha\xi}}$ independent samples from any unknown distribution $D$ on a finite domain $\calU$ that answers an interval request with \Accept, \High, or \Low, answers a numerical request with a value $a_t$, and halts immediately after the $(T+1)$-th \High or \Low answer. 
With probability at least $1-\xi$, all answers issued by the mechanism simultaneously satisfy the following:
\begin{enumerate}
\item 
If the mechanism returns \Accept, then $\EEx{X\sim D}{f_t(X)}\in[L_t-\alpha,H_t+\alpha]$.
\item 
If the mechanism returns \High, then $\EEx{X\sim D}{f_t(X)}\geq H_t-\alpha$. 
\item
If the mechanism returns \Low, then $\EEx{X\sim D}{f_t(X)}\leq L_t+\alpha$.
\item 
All answers $a_t$ satisfy $\left|a_t-\EEx{X\sim D}{f_t(X)}\right|\leq\alpha$. 
\end{enumerate}
\end{corollary}
The main point is that interval requests can be handled using a standard \textsc{AboveThreshold} mechanism, allowing arbitrarily many \Accept responses in a single phase and refreshing the threshold noise only after a \High or \Low response. 
Thus, the additional privacy composition involves at most $T+1$ threshold phases and $R$ numerical releases, rather than the total number $m$ of requests.
The mechanism's hierarchical structure remains unchanged, and ensuring accuracy simultaneously over all $m$ requests only incurs logarithmic overheads. 
This explains the $\frac{\sqrt{R+T}}{\alpha}$ term in the sample complexity, with the $\frac{1}{\alpha^2}$ term supplying the samples needed for the reference weights and internal corrections, as before.

\section{Robust \texorpdfstring{$F_2$}{F2} Algorithm}
In this section, we describe our adversarially robust $F_2$ estimation algorithm for insertion-deletion streams, which crucially relies on the improved ADA sample complexity shown in the previous section. 
Recall that in the classical streaming model, it is known that estimating the $F_2$ moment up to $(1+\eps)$-multiplicative error requires $\Omega\left(\frac{1}{\eps^2} \log n\right)$ bits of space \cite{KaneNW10a}. 
As our algorithm uses only $\tO{\frac{1}{\eps^2}}$ bits of space throughout the stream of length $m =\poly(n)$, this shows that there is essentially \textit{no separation} in the space complexity of $F_2$ estimation in the oblivious and adaptive settings, up to $\polylog(n)$ factors. 

\subsection{Overview of \texorpdfstring{\cite{GribelyukLWYZ26}}{GLWYZ26}}

Before we present our optimal $F_2$ algorithm, we briefly give an overview of the techniques developed in \cite{GribelyukLWYZ26}, which designed a $\tO{\frac{1}{\eps^4}}$ space adversarially robust $F_2$ algorithm in the turnstile setting. 
Suppose that $\bx = \bz + \bq$ is the frequency vector, and the algorithm maintains $\bA \bz$ and $\bB \bq$ for independent $F_2$ sketches $\bA, \bB$. 
Additionally, the algorithm maintains sketch $\bA \bx$ in the background. 
The algorithm is centered around three components:

\begin{enumerate}
    \item The estimator computes an approximation of $\|\bz + \bq\|_2^2$ via $\|\bz - \bv\|_2^2 + \|\bv + \bq\|_2^2$, where $\bv$ is the current iterate. Notably, the first term $\|\bz - \bv\|_2^2$ is estimated using $\bA$ and $\|\bv + \bq\|_2^2$ is estimated using sketch $\bB$.
    \item The corrector maintains sketch $\bA \bx$, and before the estimator releases an approximation of $\|\bz - \bv\|_2^2 + \|\bv + \bq\|_2^2$, the algorithm checks that the estimator's value is within $(1\pm \eps) \cdot \|\bx\|_2^2 $. If the estimator is inaccurate, the corrector releases an estimate of $\|\bx\|_2^2$ using $\bA \bx$ and flags the estimator as inaccurate.

    \item The learner maintains an iterate $\bv$, which is initialized as $\bv = \mathbf{0}$. If the corrector flags the current estimate $\|\bz - \bv\|_2^2 + \|\bv + \bq\|_2^2$ as inaccurate, the learner updates the current $\bv$ and passes the updated iterate to the estimator.
\end{enumerate}
\allowdisplaybreaks

We emphasize that all vectors are only accessible in the sketch space.
In fact, \cite{GribelyukLWYZ26} gave a strategy to update $\bv$ so that each time the estimator is not accurate, the squared distance between $\bz$ and the new iterate decreases by a factor of $(1-\eps^2)$: if $\bv$ is updated to $\bv'$, we have $\|\bz - \bv'\|_2^2 \leq (1-\eps^2) \|\bz - \bv\|_2^2$. Since $\|\bz \|_2^2 \leq \poly(n)$, this implies that $\bA$ will be used to flag the estimator as inaccurate and subsequently to release a new estimate at most $L = \O{\frac{1}{\eps^2} \log n}$ times throughout the stream. To limit the adaptive interactions of the adversary with each sketch $\bA$ and $\bB$, the key insight is that estimating $\|\bv + \bq\|_2^2$ is a task of precisely the same form as the original problem. Thus, the algorithm recursively computes $\|\bv + \bq\|_2^2 $ by subdividing $\bq = \bq_1 + \bq_2$ and using the estimator $\|\bv + \bq_1 - \bu\|_2^2 + \|\bu + \bq_2\|_2^2$, where each term is estimated using an independent sketch for $F_2$. 

More generally, the algorithm maintains a $B$-ary recursive tree of height $H = \O{\log n}$ over the stream of length $m = \poly(n)$, such that there is only a single active block in each level of the tree at a given time. Each block has $B$ child blocks and uses an independent $F_2$ sketch to maintain intermediate vectors and compute estimates. Importantly, note that the recursive computation of $\|\bv + \bq\|_2^2$ is only well-defined for a \textit{fixed} iterate $\bv$ of the parent block. To handle this, the algorithm of \cite{GribelyukLWYZ26} initializes a new block each time that the iterate of the parent block changes, or $B$ child blocks have passed, so that $B \geq L$. At a high-level, this means that each corrector sketch (and estimator sketch) must handle $\O{B \cdot L}$ adaptive updates. Then, by applying the differential-privacy robustification framework of \cite{HassidimKMMS20}, it follows that the sketch in each block uses \[\tO{\sqrt{B \cdot L}} \cdot \frac{1}{\eps^2} \log n\] bits of memory. As the full algorithm stores $H = \O{\log n}$ sketches at any given time, it follows that the total space is $\tO{\frac{1}{\eps^4}}$ bits overall.

\paragraph{Toward an optimal space algorithm.} To improve the robust $F_2$ algorithm described above, we first present an alternate recursive algorithm which allows us to maintain a \textit{binary} tree; namely, unlike the previous algorithm, we do not need to initialize a new block each time that the parent block's iterate changes. This means that by setting $B = 2$ and applying the robustification framework of \cite{HassidimKMMS20} to ensure that each sketch can handle $\O{B \cdot L} = \O{\frac{1}{\eps^2} \log n}$ adaptive queries, this modified recursive algorithm already yields an improved $\tO{\frac{1}{\eps^3}}$ space adversarially robust $F_2$ algorithm. To obtain the final $\tO{\frac{1}{\eps^2}}$ space algorithm, observe that it suffices to devise a mechanism that, given a linear sketch $\bA$ with $\tO{\frac{1}{\eps^2}}$ rows, maintains $\bA \bx$ and can be used to output $F_2$ estimates for $\tO{\frac{1}{\eps^2}}$ adaptive queries. To this end, we use our new mechanism for adaptive data analysis to circumvent the $\tO{\frac{1}{\eps^3}}$ dependence incurred by the naive DP-robustification, and obtain an algorithm with the optimal $\tO{\frac{1}{\eps^2}}$ dependence on the accuracy.

The following subsections are organized as follows. In \Cref{sec:toy:problem}, we describe a linear sketch for $F_2$ estimation which answers $\tO{\frac{1}{\eps^2}}$ adaptive queries using $\tO{\frac{1}{\eps^2}}$ rows. This subroutine will be used within each block of the recursive hierarchy described above. Then, in \Cref{sec:improved:rec} we present the full $F_2$ algorithm under adaptive insertions and deletions by combining this subroutine with a modified recursion, which maintains a binary tree of sketches over the stream.

\subsection{Optimal Linear Sketch for Adaptive Queries}\label{sec:toy:problem}
We now establish the existence of a linear sketch which uses $\tO{\frac{\sqrt{k}}{\eps} + \frac{1}{\eps^2}}$ rows and outputs a $(1\pm \eps)$-approximation to the $F_2$ moment for $k$ adaptive queries. First, we recall the classical AMS sketch of \cite{AlonMS99} which provides a $(1\pm \eps)$-approximation to the $F_2$ moment in the classical streaming setting.

\begin{theorem}[AMS Algorithm for $F_2$ estimation \cite{AlonMS99}]

\label{thm:ams}
Let $\eps\in(0,1)$ be an accuracy parameter and $\delta\in(0,1)$ be a failure probability. 
Let $r=\O{\frac{1}{\eps^2}\log\frac{1}{\delta}}$. 
There exists a turnstile streaming algorithm that maintains a linear sketch $\bA\bx$, where $\bA \in \mathbb{R}^{r \times n}$ is drawn from an explicit distribution over random matrices, uses $\O{r\log n}$ bits of space, and with probability $1-\delta$, outputs a $(1+\eps)$-approximation to the $F_2$ moment on streams of length $m=\poly(n)$. Specifically, each entry $\bA_{i,j}\sim \{1, -1\}$ is drawn from a $s$-wise independent distribution for $s = \O{\log(1/\eps)}$, and the space overhead is multiplied by $s$.
\end{theorem}

For the final estimator, the algorithm needs to draw entries $\bA_{i,j} \sim \{1, -1\}$ such that signs are $2s$-wise independent, where $s = \O{\log(1/\eps)}$. In the following lemma, we give the precise moment bounds for this variant of the AMS sketch with limited-independence.

\begin{lemma}
\label{lem:ams-moments}
Let $s \ge 1$ be an integer, and let
$r_1,\ldots,r_n \in \{-1,+1\}$ be $2s$-wise independent
uniform random signs. For any fixed vector
$x \in \mathbb{R}^n$, define
\[
Z = \sum_{i=1}^n r_i x_i,
\qquad
F = \|x\|_2^2 = \sum_{i=1}^n x_i^2.
\]
Then
\[
\mathbb{E}[Z^{2s}]
\le (2s-1)!!\,F^s
= \frac{(2s)!}{2^s s!}F^s
\le (2s)^s F^s.
\]
\end{lemma}

\begin{proof}
Expanding the $2s$-th moment gives
\[
\mathbb{E}[Z^{2s}]
=
\sum_{i_1,\ldots,i_{2s}\in[n]}
\left(\prod_{j=1}^{2s}x_{i_j}\right)
\mathbb{E}\left[\prod_{j=1}^{2s}r_{i_j}\right].
\]
For each tuple $(i_1,\ldots,i_{2s})$, let $m_i$
denote the number of occurrences of index $i$.
Since $r_i^2=1$, we have
\[
\prod_{j=1}^{2s}r_{i_j}
=
\prod_{i:\,m_i\text{ odd}}r_i.
\]
By $2s$-wise independence and the fact that each
sign has mean zero,
\[
\mathbb{E}\left[\prod_{j=1}^{2s}r_{i_j}\right]
=
\begin{cases}
1, & \text{if every }m_i\text{ is even},\\
0, & \text{otherwise}.
\end{cases}
\]
Therefore,
\begin{equation}
\label{eq:ams-even-tuples}
\mathbb{E}[Z^{2s}]
=
\sum_{\substack{i_1,\ldots,i_{2s}\in[n]\\
                 m_i\text{ even for all }i}}
\prod_{j=1}^{2s}x_{i_j}.
\end{equation}

We bound this sum by counting pairings. A pairing of
$[2s]$ is a partition of its elements into $s$
unordered pairs. The number of such pairings is
\[
|\mathcal{P}_{2s}|
=(2s-1)(2s-3)\cdots 3\cdot 1
=\frac{(2s)!}{2^s s!}.
\]

Every tuple appearing in
\eqref{eq:ams-even-tuples} is compatible with at
least one pairing: since each index occurs an even
number of times, its occurrences can be paired
among themselves.

Fix a pairing $\pi\in\mathcal{P}_{2s}$. Summing over
all tuples whose indices agree within each pair gives
\[
\sum_{\substack{i_1,\ldots,i_{2s}\in[n]\\
                 i_a=i_b\text{ for every }(a,b)\in\pi}}
\prod_{j=1}^{2s}x_{i_j}
=
\sum_{j_1,\ldots,j_s\in[n]}
\prod_{\ell=1}^s x_{j_\ell}^2
=
\left(\sum_{i=1}^n x_i^2\right)^s
=F^s.
\]

Every summand in \eqref{eq:ams-even-tuples} is
nonnegative, because all indices occur with even
multiplicity. Thus summing over all pairings can
only overcount the surviving terms. It follows that
\[
\mathbb{E}[Z^{2s}]
\le \sum_{\pi\in\mathcal{P}_{2s}}F^s
=\frac{(2s)!}{2^s s!}F^s
\le (2s)^sF^s.
\]

\end{proof}

\paragraph{Adaptive query model.}

We consider the adaptive query model for linear sketching, defined as follows.

\begin{enumerate}
    \item At each time $t \in [k]$, the adversary selects a query vector $\bx_t \in \mathbb{Z}^n$, which may depend on the sequence of previous queries $\{\bx_i\}_{i < t}$ and previous outputs of the sketch $\{Z_i\}_{i < t}$. 
    \item The sketching algorithm observes $\bA \bx_{t}$ and returns an estimate $Z_t$ for $\|\bx_t\|_2^2$. 
\end{enumerate}

We remark that an algorithm for the adaptive query model can be implemented in the streaming model by feeding turnstile updates $u_t = (a_t, \Delta_t)$ defining an adaptive query $\bx$ to the algorithm sequentially, under the promise that the sketch will output an estimate at most $k$ times throughout the stream. 

In fact, our algorithm for answering adaptive $F_2$ estimation queries using a linear sketch crucially relies on our improved adaptive data analysis mechanism for statistical queries, as described in the previous section. Concretely, suppose $\bA$ is initialized as an independent AMS sketch with $2s$-wise independent signs using \Cref{thm:ams}. Then, for a given query vector $\bx_t \in \mathbb{Z}^n$, the standard AMS algorithm computes 
\[\frac{1}{r}\|\bA \bx_t\|_2^2 = \frac{1}{r} \sum_{i = 1}^r \langle \bA_i, \bx_t \rangle^2   .\]

While this resembles a statistical query, the main issue is that $\langle \bA_i, \bx_t \rangle^2$ is not necessarily in the interval $[0, 1]$. To handle this, we initialize independent AMS sketches $\bB_t$ with $\O{\log n \log(k/\delta)}$ rows for each $t \in [k]$, such that $\frac{1}{2} \|\bx_t \|_2^2 \leq \|\bB_t\bx_t\|_2^2 \leq 2 \|\bx_t\|_2^2$ with probability $1-\delta/(2k)$. Then, we define the $t$-th statistical query by normalizing,
\[f_t(\by) = \min\left(1, \frac{\langle \by, \bx_t \rangle^2 }{L \cdot \|\bB_t \bx_t\|_2^2} \right),\]
where $L= 16 s$. For the case that $\|\bB_t \bx_t\|_2^2 = 0$ we define $f_t(\by) = 0$. The final algorithm is summarized below.

\begin{figure*}[!htb]
\begin{mdframed}
\textbf{Algorithm}:
\begin{enumerate}
\item Initialize a $2s$-wise independent AMS sketch $\bA$ with $r = \tO{\frac{\sqrt{k}}{\alpha} + \frac{1}{\alpha^2}}$ rows, with accuracy parameter $\alpha = \frac{\eps}{64s}$ for $s = \O{\log(1/\eps)}$, and failure probability $\rho = \delta/4$. For each $t\in [k]$, let $\bB_t$ be a $4$-wise independent AMS sketch with $\O{\log n \cdot \log(k/\delta)}$ rows.

\item Initialize an instance $\calA$ of the ADA mechanism, with sample $S:= \{\bA_i\}_{i \in [r]}$ of rows of $\bA$, where $r = \tO{\frac{\sqrt{k}}{\alpha} + \frac{1}{\alpha^2}}$. 
\item For each query vector $\bx_t \in \mathbb{Z}^n$ for $t \in [k]$:
\begin{itemize}
    \item Compute estimate $b_t \gets \|\bB_t \bx_t\|_2^2$. Let $f_t(\by) = \min\left(1, \frac{\langle \by, \bx_t \rangle^2 }{L \cdot b_t} \right)$ be the capped query. 
    \item Feed query $\bar f_t(\by) =   \frac{\lfloor D \cdot f_t(\by) \rfloor }{D}$ to $\calA$ for $D = \O{\frac{1}{\alpha}}$ and obtain estimate $a_t \gets \calA(\bar f_t)$.
    \item Return $L \cdot a_t \cdot b_t$.
    
\end{itemize}

\end{enumerate}
\end{mdframed}
\caption{Adversarially robust sketch for $F_2$ queries.}
\label{fig:alg:ltwo:opt}
\end{figure*}

\paragraph{Algorithm analysis.} 
We now verify the correctness guarantees of our sketch.
\begin{lemma}
For any $\bx_t \in \mathbb{Z}^n$, let $\frac{1}{2} \|\bx_t\|_2^2\leq b_t \leq 2\|\bx_t\|_2^2$ be the estimate produced by sketch $\bB_t$, for each $t \in [k]$. Then, the query $\bar f_t$ satisfies
\[|\mathbb{E}_{\bu}[\bar f(\bu)] - \mathbb{E}_{\bu}[f(\bu)]| \leq \alpha/4\]
and 
\[0 \leq \|\bx_t\|_2^2 - b_t L \cdot \mathbb{E}_{\bu}[f(\bu)] \leq \frac{\eps \|\bx_t\|_2^2}{8},\]
where each entry of $\bu$ is drawn from a $2s$-wise independent distribution over $\{1, -1\}$.
\end{lemma}
\begin{proof}
Let $X = \langle \bu, \bx_t \rangle$ where $\bu \sim \{1, -1\}^n$, where each entry is $2s$-wise independent. Conditioned on the correctness of sketch $\bB_t$, we have that $b_t \geq \frac{1}{2} \cdot \|\bx_t\|_2^2$, so by \Cref{lem:ams-moments} we have
\begin{align*}
\|\bx_t\|_2^2 - b_t L \cdot \mathbb{E}_{\bu}[f(\bu)] &= \mathbb{E}_{\bu}\left[(X^2 - L \cdot b_t)_{+} \right] \\ & \leq \frac{\mathbb{E}[X^{2s}]}{(L \cdot b_t)^{s-1}} \\ & \le \frac{(2s)^s \|\bx_t\|_2^{2s}}{(L\|\bx_t\|_2^2/2)^{s-1}} \\ &= 2s \left(\frac{4s}{L} \right)^{s-1} \|\bx_t\|_2^2 = 2s \cdot 4^{-s+1} \|\bx_t\|_2^2.
\end{align*}
Note that $2s \leq 2^s$, so it follows that the difference $\|\bx_t\|_2^2 - b_t L \cdot \mathbb{E}_{\bu} [f_t(\bu)] \leq \frac{\eps}{8} \|\bx_t\|_2^2$ overall. Moreover, note that the difference between $f_t$ and $\bar f_t$ satisfies 
\[0 \leq f_t(\bu)- \bar f_t(\bu) \leq 1/D\]
for each input, so the claim follows.
\end{proof}

\begin{observation}
For each $t \in [k]$, $\bar f_t$ represents a valid adaptive statistical query to the ADA mechanism $\calA$, and with high probability,
\[|a_t - \mathbb{E}_{\bu} \bar f_t(\bu)| \leq \alpha.\]
\end{observation}
In particular, note that at each time $t$, the linear sketch observes sketches $\bA \bx_t$ and $\bB_t \bx_t$ (where  $\bx_t$ is adaptively chosen), and forms the statistical query $\bar f_t(\cdot)$, which is fed to the ADA mechanism. 

\begin{lemma}[Adversarially robust sketch]\label{lem:robust:sketch}
Let $\eps, \delta \in (0,1)$. There is an estimator in the adaptive query model such that, for every $t \in [k]$ simultaneously, with probability at least $1-\delta$, the estimator's outputs $Z_t$ satisfies
\[(1-\eps) \cdot \|\bx_t\|_2^2 \leq Z_t \leq (1+\eps) \cdot \|\bx_t\|_2^2 .\]
The sketch matrix has most $r \leq \O{\left(k + \frac{\sqrt{k}}{\eps} + \frac{1}{\eps^2}\right) \cdot \polylog \frac{1}{\eps \delta}}$ rows.
\end{lemma} 
\begin{proof}
We first show the accuracy guarantee of our algorithm. Conditioned on the correctness of the $\bA$ and $\bB_t$ sketches for all $t \in [k]$, and recalling that $\alpha = \frac{\eps}{64s}$, it follows that
\begin{align*}
|\hat F_t - F_t| &\leq L b_t |a_t - \mathbb{E} [\bar f_t] | + L b_t |\mathbb{E}[\bar f_t] - \mathbb{E}[f_t]| + |L b_t \mathbb{E}[f_t] - F_t| \\ & \leq L b_t (\alpha + \frac{\alpha}{4}) + \frac{\eps}{8} F_t \\ &\leq \eps \cdot F_t.
\end{align*}
As the $k$ independent sketches $\bB_t$ provide a $2$-approximation to the squared norm $\|\bx_t\|_2^2$ with probability $1-\delta/(4k)$, by a union bound it follows that all sketches $\bB_t$ provide accurate estimates with probability $1-\delta/4$. Additionally, the ADA mechanism $\calA$ outputs all $k$ answers satisfying $\left  | a_t - \mathbb{E}[\bar f_t(\by)] \right | \leq \alpha$ with probability $1-\delta/4$, so the overall sketch is $(1+\eps)$-approximate at all $k$ times with probability at least $1-\delta$. 

Next, we consider the space complexity of the algorithm. Recall that $s = \O{\log(1/\eps)}$ and $\alpha = \O{\frac{\eps}{s}}$. The algorithm maintains $k$ sketches $\bB_t$ for $t \in [k]$, each of which uses $\O{\log^2 n \log(k/\delta)}$ bits of space. Also by construction and \Cref{thm:ada:main}, $\bA$ has $\tO{\frac{\sqrt{k}}{\alpha} + \frac{1}{\alpha^2}} = \tO{\frac{\sqrt{k}}{\eps} \log \frac{1}{\eps} + \frac{1}{\eps^2} \log^2 \frac{1}{\eps}}$ rows which are generated by a $2s$-wise independent hash function, so the total space is 
\[\tO{\left(\frac{\sqrt{k}}{\eps} \log \left(\frac{1}{\eps}\right) + \frac{1}{\eps^2} \log^2 \left(\frac{1}{\eps}\right)\right)\log \left(\frac{1}{\delta}\right) \log n  + k \log^2 n \log(k/\delta)}\]
bits of memory for storing $\bA$ and all sketches $\bB_t$ for $t \in [k]$, and the claimed space complexity follows.
\end{proof}

\begin{remark}\label{rem:cond:rel:linear}
The adaptive conditional release task in  \Cref{cor:conditional:release} with $m$ queries, $R$ numerical requests, and $R$ \High or \Low outputs can be solved using a linear sketch and estimator with $\tO{R + \frac{\sqrt{R}}{\eps} + \frac{1}{\eps^2}}$ rows. Concretely, the algorithm stores $\O{R}$ $2$-approximation sketches $\bB_t$ for $t \in [R]$ which are used to determine the scale of the estimate, and a new sketch $\bB_{t+1}$ is used once $\bB_t$ has been used during a  \High or \Low outcome followed by a numerical release. 
\end{remark}

\subsection{An Improved Recursive Algorithm}
\label{sec:improved:rec}
We now present our modified recursive algorithm for $F_2$ moment estimation which enables the final nearly-optimal $\tO{\frac{1}{\eps^2}}$ space complexity. In order to reduce the dependence on $\frac{1}{\eps}$ in the prior robust $F_2$ algorithm \cite{GribelyukLWYZ26}, we must ensure that each sketch in the hierarchy only needs to serve as a corrector for $\O{1}$ child blocks. To this end, we reframe the estimator-corrector-learner framework to enable a \textit{binary} tree of sketches over the stream.

\begin{figure*}[!htb]
\begin{mdframed}
\textbf{Algorithm}:
\begin{enumerate}
\item 
Maintain a tree structure over the stream of length $m = \poly(n)$. The bottom level $1$, each block contains a single update.
\item
Each new node in level $i$ is formed when two blocks in level $i-1$ have passed in the stream.
\item 
For each block $C_{i,j}$, use sketch matrix $\bB_{i,j}$ to sketch all necessary vectors, such as the frequency vector in the block and the iterate vectors, as required by the subroutines.
\item
Let $H$ be the height of the tree and let $\bB_{H+1}$ be a sketch matrix at the top.
\item
Output $\EstLevel(H+1,\bB_{H+1}\bz)$, where $\bz$ is the frequency vector for the stream.
\end{enumerate}
\end{mdframed}
\caption{Adversarially robust $F_2$ norm estimation algorithm on insertion-deletion streams}
\label{fig:alg:ltwo:improv}
\end{figure*}

\paragraph{Reformulated update rule.} For now, let us focus on the following simplified setting: suppose the frequency vector is $\bx = \bz + \bq$ as before, and $\bA$ and $\bB$ are independent $F_2$ sketches via \Cref{lem:robust:sketch}. Furthermore, suppose the algorithm maintains $\bA \bz$ and $\bB \bq$, as well as an iterate $\bv$ (maintained using both $\bA$ and $\bB$) which approximates the prefix $\bz$. Recall that the estimator computes
\[\| \bz - \bv\|_2^2 + \|\bv + \bq \|_2^2\]
and the corrector sketch maintains an estimate of $\|\bz + \bq\|_2^2$. Suppose the estimator is not accurate for the current suffix $\bq$; then, the learner updates $\bv$ to $\bv'$, which is defined by $\bv' = \bv + \alpha(\bv + \bq)$ for step-size parameter $|\alpha| = \frac{\eps \|\bz - \bv\|_2}{\|\bv + \bq\|_2}$, where the sign is given by $\textrm{sign}(\langle \bz - \bv, \bv + \bq \rangle )$. Let $\bv_0 = \mathbf{0}$ be the iterate initially, and let $\bq_0$ be the first suffix for which the estimator above is inaccurate. This implies
\[\bv_1 = \bv_0 + \alpha_1(\bv_0 + \bq_0),\]
so the sketch $\bB$ will subsequently be used to sketch $\bB(\bv_1 + \bq_0) = (1+\alpha_1) \cdot \bB(\bv_0 + \bq_0)$. Let $\gamma_i = 1+\alpha_i$ denote this scalar factor at the $i$-th update. 
Now, suppose that $\bq_1$ represents the vector of \textit{new} updates received after the first flagged suffix $\bq_0$; as before, the new iterate is defined as
\[\bv_2 = \bv_1 + \alpha_2(\bv_1 + \bq_0 + \bq_1),\]
and the sketch will be updated to maintain $\bB(\bv_2 + \bq_0 + \bq_1) = \gamma_2 \cdot \bB(\bv_1 + \bq_0 + \bq_1)$. For $\bv_1$ defined as above, the sketch must compute $\gamma_1 \gamma_2 \cdot \bB(\bv_0 + \bq_0 + \frac{\bq_1}{\gamma_1})$. It follows that at the $k$-th step, the input to sketch $\bB$ can be expressed as 
\[\left(\prod_{i \in [k]} \gamma_i \right) \bB\left(\bv_0 + \bq_0 + \frac{\bq_1}{\gamma_1} + \ldots + \frac{\bq_k}{\gamma_1 \cdots \gamma_k}\right).\]

\paragraph{\EstLevel and \MaintainIter subroutines.}
For the complete algorithm, we maintain a binary tree of sketches over the stream of length $m$, such that a single block $\calP_i$ is active at each level of the tree at any given time. The $\EstLevel(i, \bB_i \bz)$ subroutine then recursively computes an estimate of the current $F_2$ moment of its input substream $\bx = \bz + \bq$ by estimating $P_i$ using $\bB_i$ and computing $Q_i$ recursively via $\EstLevel(i-1, \bB_{i-1} (\bv + \bq))$. Before returning $P_i + Q_i$, the estimator compares its estimate to the value of the corrector sketch $A_i$, and invokes $\MaintainIter(i)$ if the estimate is not sufficiently accurate. Importantly, $\MaintainIter(i)$ propagates the scalar multiplier $(1+\alpha)$ to its child blocks but \textit{re-uses} cached estimates for $P_j$ for $j < i$, which are simply multiplied by $(1+\alpha)^2$ when an estimate is requested from the lower levels. For instance, suppose that the child blocks partition this frequency vector $\by = \bv_k + \sum_{i = 0}^{k} \bq_i$ such that the left child maintains $\bB\bu$ and the right child maintains $\bC\bw$, where $\bu + \bw = \by$ and $\bC$ is an independent $F_2$ sketch. Then, when the iterate $\bv_k$ is updated to $\bv_{k+1}$ in the parent node, we simply multiply the sketched values $\bB \bu$ and $\bC \bw$ by $\gamma_{k+1}$, and multiply the cached estimate of $\|\bu\|_2^2$ by $\gamma_{k+1}^2$. By factoring out the product $\left(\prod_{i \in [k+1]} \gamma_i \right)$ from the sketched vector, we note that this multiplicative scaling does not interfere with the progress already made by the learner in the child node.

\begin{algorithm}[!htb]
\caption{$\EstLevel(i,\bB_i\bz)$ for $F_2$ moment estimation}
\label{alg:est:level:ltwo}
\begin{algorithmic}[1]
\State{Let $\calP_i$ be the active block at level $i$}
\State{Let $\bB_i$ be the sketch matrix for $\calP_i$}
\State{}\Comment{$\left(1+\eta\right)$-approximation for $F_2$ moment  for $\eta = \frac{\eps}{100H}$ via \Cref{lem:robust:sketch}, robust to $\O{\frac{1}{\eta^2} \log n}$ adaptive queries}
\State{Let $A_i$ be a $\left(1+\eta\right)$-approximation to $\|\bz+\bq\|_2^2$}
\If{$i\neq 1$}
\State{$P_i\gets\bB_i(\bz +\bq_0 - \bv')$}

\State{}\Comment{$\bv'$ is defined as $\bv' := \bv + \bw$, where $\bw$ contains all additional coordinate updates.}
\State{$Q_i\gets\EstLevel(i-1,\bB_{i-1}(\bv'+\bq_1))$}
\Else
\State{$P_i\gets 0$}
\State{Let $\bq_1$ be the active part of the query in $\calP_i$}
\State{$Q_i\gets\bB_1(\bv' + \bq_0 + \bq_1)$}
\EndIf
\If{$P_i+Q_i\in\left(1\pm\eta\right)^{3i}\cdot A_i$}
\State{\Return $P_i + Q_i$}
\Else
\State{$\bv \gets \MaintainIter(i)$}
\State{\Return $A_i$}
\EndIf
\end{algorithmic}
\end{algorithm}

\begin{algorithm}[!htb]
\caption{$\MaintainIter(i)$ for $F_2$ moment estimation}
\label{alg:mainiter:ltwo}
\begin{algorithmic}[1]
\State{Initialize $\bv=\textbf{0}^n$ at the beginning of the active block at level $i$}
\State{$\bq\gets\bq_0+\bq_1$}
\If{$P_i+Q_i\notin\left(1\pm\frac{\eps}{100H}\right)^{3i}\cdot A_i$} \Comment{Implement interval check via \Cref{cor:conditional:release}}
\State{$\sigma \gets \textrm{sign}(A_i - (P_i + Q_i))$}
\State{$\alpha\gets\frac{|A_i - (P_i + Q_i)|}{2Q_i}\cdot\sigma$} \Comment{Step-size is the minimizer of regularized optimization.}
\State{$\bv\gets\bv+\alpha(\bv+\bq)$, $\bw \gets (1+\alpha) \bw $} 
\State{Update all child blocks: re-use cached estimates for $P_{j}$ for $j < i$ and multiply by $(1+\alpha)^2$}
\EndIf
\State{\Return $\bv$ truncated to precision $\frac{1}{\poly(n)}$}
\Comment{All operations performed in the sketch space}
\end{algorithmic}
\end{algorithm}

\paragraph{Bounded bit complexity.}
While the description above allows us to \textit{avoid restarting} the estimator-learner-corrector framework at child sketches $\bB$ and $\bC$ each time that the parent node's iterate is updated, it remains to ensure that the bit complexity of values stored by sketches can be bounded by $\polylog(n)$. Naively, it may seem that the product $\prod_{i \in [k]} \gamma_i$ may be as large as $2^{\O{\frac{1}{\eps^2} \log n}}$, requiring us to track $\O{\frac{1}{\eps^2} \log n}$ bits of precision in each of the $r = \tO{\frac{1}{\eps^2}}$ dimensions of the sketched vector. In the following, we show a way to modify our algorithm to enable storing just $\polylog(n)$ bits of precision, while still achieving the desired $(1+\eps)$-multiplicative error.

Consider the vectors $\bz - \bv_t$ and $\bv_t + \bq$ sketched by $\bA$ and $\bB$ after the $t$-th update to the iterate $\bv_t$, and suppose $\|\bz - \bv_t\|_2^2 \leq M$ and that $\|\bz - \bv_s\|_2^2 \leq (1-\eps^2) \|\bz -\bv_{s-1}\|_2^2$ for all $s \leq t$. As before, let $L = \O{\frac{1}{\eps^2} \log n}$ denote the maximum number of updates to the iterate. To overcome the precision issue described earlier, we instead imagine that before each query, the vectors sketched by $\bA$ and $\bB$ are extended to have a larger universe of size $\poly(n)$, and add $\delta \cdot e_{\ell}$ to $\bv$ to obtain sketches of $\bz - \bv_t'$ and $\bv_t' + \bq$, where $\ell > n$ is a previously unused index of the frequency vector and $\delta = \frac{1}{\poly(n)}\sqrt{M}$ for some small enough $\frac{1}{\poly(n)}$; we show that these modified frequency vectors admit a $\O{\log n}$-bit representation. In the following analysis and algorithm description, we instead implement the above argument as solving a \textit{regularized} optimization problem to compute the appropriate step-size $\alpha_t$ in each step $t \in [L]$ which (1) ensures that $P_{s,t} \leq \poly(n)$ for all $s$ and $t$, and (2) the learner makes sufficient progress toward $\bz$. 

\paragraph{Regularized optimization.}
Let $\bz$ be the prefix, $\bv_t$ be the $t$-th iterate, and let $\bw$ be the vector containing $\delta$-scaled coordinates. Then, the squared norm of the imaginary sketched vector $\bz - \bv_t + \bw$ is precisely $\|\bz - \bv_t\|_2^2 + \|\bw\|_2^2$. Now, suppose we insert a new $\delta$-coordinate in an unused index in $\bw$. For current suffix $\bq$, the iterate is updated as $\bv_{t} = \bv_{t-1} + \alpha_{t} (\bv_{t-1} + \bq)$. To compute the chosen step-size $\alpha_t$, we consider the objective 

\[\Psi(\alpha_t) = \|\bz - \bv - \alpha_t(\bv + \bq)\|_2^2 + (1+\alpha_t)^2 (\|\bw\|_2^2 ).\]

\begin{lemma}\label{lem:reg:step:size}
    Consider the minimization problem $\alpha_t = \argmin_{\alpha \in \mathbb{R}}\Psi(\alpha)$. The minimizer is given by \[\alpha_t = \frac{\langle \bz -\bv_t, \bv_t + \bq \rangle - \|\bw\|_2^2 }{\|\bv_t + \bq\|_2^2 + \|\bw\|_2^2}.\]
\end{lemma}

\begin{proof}
Note that 
\[\Psi(\alpha_t) = \alpha_t^2 \left(\|\bw\|_2^2 + \|\bv_t + \bq\|_2^2 \right) + 2\alpha_t \left(\|\bw\|_2^2 - \langle \bz - \bv_t, \bv_t + \bq\rangle \right) + \|\bz - \bv_t\|_2^2 + \|\bw\|_2^2.\]
By viewing the above as a quadratic optimization with respect to $\alpha$, the unique minimizer is as claimed.
\end{proof}

The remaining argument proceeds by induction: suppose that $\|\bz - \bv_s'\|_2^2 \leq (1-\eta^2) \cdot \|\bz - \bv_{s-1}'\|_2^2$ at all times $s \leq t$, where $\eta = \O{\frac{\eps}{H}}$. 
First, we will argue that the bit-complexity of various scalars maintained by the algorithm is bounded; 
Using this, we will then conclude that the learner's new update will shift the modified iterate $\bv_{t+1}'$ closer to $\bz$.
\begin{lemma}[Bounded Products]\label{lem:bit:complexity}
Let $M = \poly(m) $ be a uniform upper bound on the maximum possible value of $\|\bz\|_2^2$, and let $\calP_i$ be the active block in level $i \in [H]$. 
Let $\gamma_j = 1+\alpha_j$ denote the scalar multiplying all sketched vectors in the $j$-th iterate vector $\bv_j$ in $\calP_i$. 
Suppose that $\|\bz - \bv_s'\|_2^2 \leq (1-\eta^2) \cdot \|\bz - \bv_{s-1}'\|_2^2$ is satisfied at all times $s \leq t$. Then, for each $s \leq t$, we have $\left | \prod_{j = s}^t \gamma_j \right | \leq M^{5/2}$.
\end{lemma}
\begin{proof}
Let $\bv_t'$ be the iterate with the added $\delta$-updates in previously unused coordinates (which will only be used at most once), and let $P_{s,t} = \prod_{j = s}^t \gamma_j$. 
By assumption, it holds that $\|\bz - \bv_s'\|_2^2 \leq M$ at all update times $s \leq t$. 
Since we inject a new coordinate in an unused index before each query, there is an added $\delta$-coordinate that will be multiplied by $P_{s,t}$, and we have
\[\delta^2 P_{s, t}^2 \leq \|\bz - \bv_t'\|_2^2 \leq M.\]
In particular, this implies that $|P_{s,t}| \leq \frac{\sqrt{M}}{\delta}$. By setting $\delta = \Theta\left(\frac{1}{M^2}\right)$ it follows that $|P_{s,t}| \leq M^{5/2}$. 
\end{proof}

In the next few lemmas, we show the accuracy guarantees of our algorithm.

\begin{lemma}\label{lem:p:term}
Let $\bv$ be a fixed iterate in active block $\calP_i$ for $i \in [H]$. Suppose $P_i$ is computed using sketch $\bB_i$, which is robust for $\O{\frac{1}{\eta^2}\log n}$ adaptive updates. With high probability,
\[(1-\eta) \cdot \left(\|\bz - \bv\|_2^2 + \|\bw\|_2^2 \right)\leq P_i \leq (1+\eta) \cdot \left(\|\bz - \bv\|_2^2 + \|\bw\|_2^2\right).\]
\end{lemma}
\begin{proof}
Recall that $\bB_i$ is an $(1+\eta)$-approximation sketch for $F_2$ estimation designed via \Cref{lem:robust:sketch}, which is equipped to answer $k = \O{\frac{1}{\eta^2} \log n}$ adaptive queries. Also, recall that the input to $\bB_i$ is $\bz - \bv'$, where $\bv' = \bv + \bw$ for added-coordinates $\bw$. Since $\langle \bw, \bv\rangle = 0$ and $\bB_i$ produces a $(1+\eta)$-approximation to the input, we have
\[(1-\eta) \cdot \left(\|\bz - \bv\|_2^2 + \|\bw\|_2^2 \right)\leq P_i \leq (1+\eta) \cdot \left(\|\bz - \bv\|_2^2 + \|\bw\|_2^2\right),\]
as desired.
\end{proof}

\begin{lemma}\label{lem:F2:accuracy}
Let $\bv$ be a fixed iterate in active block $\calP_i$ for $i \in [H]$. Suppose $P_i$ is computed using sketch $\bB_i$, which is robust for $\O{\frac{1}{\eta^2}\log n}$ adaptive updates, and $Q_i$ is computed by $\EstLevel(i-1, \bB_{i-1} (\bv + \bq))$ recursively. With high probability,
\[(1-\eta) \cdot \left(\|\bz - \bv\|_2^2 + \|\bv + \bq\|_2^2 + 2\|\bw\|_2^2 \right)\leq P_i + Q_i \leq (1+\eta) \cdot \left(\|\bz - \bv\|_2^2 + \|\bv + \bq\|_2^2 + 2\|\bw\|_2^2\right).\]
\end{lemma}
\begin{proof}
We proceed by induction on the level $i \in [H]$. For $i = 1$, observe that $P_1 = 0$ and $Q_1 = \bB_1(\bv' + \bq_0 + \bq_1)$, so 
\[P_1 + Q_1 \in (1\pm \eta) \cdot \left(\|\bv + \bq_0 + \bq_1\|_2^2 + \|\bw\|_2^2\right),\]
where $\bw$ is the vector of added small coordinate values at level $1$. Now, suppose by induction that the hypothesis holds for level $i-1$, meaning $Q_i = \EstLevel(i-1, \bB_{i-1}\bz)$ produces an estimate of $\|\bv' + \bq_0+ \bq_1\|_2^2$ up to multiplicative error $(1+\eta)$. In level $i$, recall that we have
\[ (1-\eta) \|\bz - \bv'\|_2^2 \leq P_i \leq (1+\eta) \|\bz - \bv'\|_2^2\]
by \Cref{lem:p:term}, conditioned on the accuracy of sketches. By union bounding over all levels $i \in [H]$, the induction step follows.
\end{proof}

As in the original robust $F_2$ algorithm \cite{GribelyukLWYZ26}, our framework ensures that the following invariant holds throughout the execution of the algorithm, at each level $i \in [H]$. 
\begin{invariant}
For level $i\in[H]$ let $\calP_i$ denote the active block in level $i$. Let $\bq$ be the part of the query which is tracked in $\calP_i$ and let $\bz$ be the prefix. Then, with high probability
\[(1-\eta)^{3i+1} \cdot \|\bz + \bq\|_2^2 \leq \EstLevel(i, \bB_i \bz) \leq \left(1 + \eta\right)^{3i+1}\cdot\|\bz + \bq\|_2^2.\]
\end{invariant}
Finally, we show that the total number of updates to the iterate is upper bounded by $\O{\frac{1}{\eta^2} \log n}$ for $\eta = \O{\frac{\eps}{H}}$. 
Specifically, we argue that the potential satisfies $\|\bz - \bv_{t+1}'\|_2^2 \leq (1-\Omega(\eta^2)) \cdot \|\bz - \bv_t'\|_2^2$, thus completing the induction defined at the beginning of \Cref{lem:bit:complexity}. 
Importantly, we note that the algorithm $Q_i = \EstLevel(i, \bB_i (\bv + \bq_0 + \bq_1))$ returns an estimate for $\|\bv + \bq\|_2^2$ by re-using the \textit{cached} estimate for $\|\bv + \bq_0 - \bu\|_2^2$; if the parent node's iterate is updated, this cached value is simply multiplied by scalars induced by the higher levels, thus ensuring that the progress at level $i$ does not re-start each time that the iterate at the parent node is updated. 

\begin{lemma}[Bounded Updates]\label{lem:bounded:updates}
Let $\calP_i$ be the active block at level $i \in [H]$, and let $L_i$ be the number of times that the iterate in $\calP_i$ is updated. With high probability, $L_i \leq \O{\frac{1}{\eta^2} \log n}$ for $\eta = \O{\frac{\eps}{H}}$ for all $i \in [H]$ over the entire stream of length $m = \poly(n)$.
\end{lemma}
\begin{proof}
Consider a fixed level $i\in [H]$, and let $\calP_i$ and $\calP_{i+1}$ represent the active blocks at levels $i$ and $i+1$. Let $\bw$ be the vector of added coordinates at level $i$. Conditioned on the accuracy guarantees of \EstLevel in \Cref{lem:F2:accuracy}, it holds that
\[(1-\eta) \cdot \left(\|\bz - \bv\|_2^2 + \|\bv + \bq\|_2^2 + 2\|\bw\|_2^2\right) \leq \EstLevel(i, \bB_{i+1}\bz) \leq (1+\eta) \cdot \left(\|\bz - \bv\|_2^2 + \|\bv + \bq\|_2^2 +2\|\bw\|_2^2\right).\]
Now, if the current iterate $\bv$ is updated, it must be that 
\[\left |  \|\bz - \bv\|_2^2 + \|\bv + \bq\|_2^2 + 2\|\bw\|_2^2 - \|\bz + \bq\|_2^2 \right| \geq \eta \cdot \|\bz + \bq\|_2^2.\]
Recall that we add auxiliary tiny $\delta$-coordinates with $\delta = \Theta\left(\frac{1}{M^2}\right)$ at each time $t$, where $M$ is a uniform upper bound on the possible value of the $F_2$ at any time. 
This means that over the entire stream, the contribution of these extra coordinates will be at most $m \log m \cdot \Theta\left(\frac{1}{M^2}\right) < \frac{1}{\poly(M)}$, so the progress analysis is unaffected by these additional coordinates. Thus, we analyze the progress assuming that $\bv'$ is the iterate vector \textit{without} the new $\delta$-valued coordinates which were added after the last update to $\bv'$. Now, recalling that $\bv'$ contains added coordinates from before the last update (which appear in $\bw$), this means that
\begin{align*}
|\langle \bz - \bv', \bv' + \bq \rangle | &\geq \frac{\eta}{2} \|\bz + \bq\|_2^2  \\ &= \frac{\eta}{2}\left(\|\bz - \bv'\|_2^2 + \|\bv' + \bq\|_2^2 + 2 \langle \bz -\bv', \bv' + \bq \rangle \right).
\end{align*}
Suppose that $\langle \bz - \bv', \bv' + \bq \rangle \geq \frac{\eta}{2} \|\bz + \bq\|_2^2$; as we show later, the other case is symmetric. 
By rearranging the inequality above, we have that
\begin{align*}
\langle \bz - \bv', \bv' + \bq \rangle &\geq \frac{\eta}{2} \left( \|\bz - \bv'\|_2^2 + \|\bv'+ \bq\|_2^2 \right) \\ & \geq  \eta \|\bz - \bv'\|_2 \|\bv' + \bq\|_2.
\end{align*}
Finally, let $\bu$ denote the updated iterate which is defined as $\bu = \bv + \alpha(\bv + \bq)$, for $\alpha = \O{1} \cdot \frac{\langle \bz -\bv', \bv' + \bq \rangle }{\|\bv' + \bq\|_2^2 }$ for any constant in the range $\left[\frac{1}{10}, \frac{1}{4} \right]$. Observe that $\frac{\langle \bz - \bv', \bv' + \bq \rangle }{\|\bv' + \bq\|_2^2} \geq \frac{\eta \|\bz - \bv'\|_2}{\|\bv' + \bq\|_2}$. 
    
Thus, we have
\begin{align*}
\|\bz - \bu'\|_2^2 &= \|\bz - \bv' - \alpha(\bv' + \bq)\|_2^2  \\ & \leq \|\bz - \bv'\|_2^2 - 2\alpha \langle \bz - \bv', \bv' + \bq \rangle + \alpha^2 \|\bv' + \bq\|_2^2 \\ & \leq \|\bz - \bv'\|_2^2 - \Omega(\eta^2) \|\bz - \bv'\|_2^2.
\end{align*}
Next, consider the remaining case when 
\[\langle \bz- \bv', \bv' + \bq \rangle  \leq - \frac{\eta}{2} \|\bz + \bq\|_2^2.\]
By expanding the right side, we have that 
\[\langle \bz - \bv', \bv' + \bq \rangle \leq -\frac{\eta}{2} \left(\|\bz - \bv'\|_2^2 + 2\langle \bz-\bv', \bv' + \bq \rangle + \|\bv' + \bq\|_2^2 \right),\]
and after combining terms, we see that 
\[- \langle \bz- \bv', \bv' + \bq \rangle \geq \frac{\eta}{4} \|\bz - \bv'\|_2 \|\bv' + \bq\|_2 .\]
As before, let $\alpha = \O{1} \cdot  \frac{\langle \bz -\bv', \bv' + \bq \rangle }{\|\bv' + \bq\|_2^2}$ be the step-size with any constant in the range $\left[\frac{1}{10}, \frac{1}{4} \right]$, and observe $\frac{\langle \bz - \bv', \bv' + \bq \rangle}{\|\bv' + \bq\|_2^2} \leq \frac{-\eta \|\bz - \bv'\|_2}{4 \|\bv' + \bq\|_2}$. Thus, it follows that
\begin{align*}
\|\bz - \bu'\|_2^2 &= \|\bz - \bv' - \alpha(\bv' + \bq)\|_2^2 \\ & \leq \|\bz - \bv'\|_2^2 -2\alpha \langle \bz - \bv', \bv' + \bq \rangle + \alpha^2 \|\bv' + \bq\|_2^2 \\ &\leq \|\bz - \bv'\|_2^2 - \Omega(\eta^2) \cdot \|\bz - \bv'\|_2^2. 
\end{align*}
So, we have shown that the potential $\|\bz - \bu'\|_2^2$ with \textit{extra coordinates} drops by a multiplicative factor of $(1-\Omega(\eta^2))$. As $\|\bz \|_2^2 \leq \poly(n)$ initially, it follows that $L_i \leq \O{\frac{1}{\eta^2} \log n}$, as desired. 
\end{proof}

\begin{theorem}[Robust $F_2$ Estimation]
For any $\eps \in (0,1)$, there exists an adversarially robust insertion-deletion streaming algorithm on a stream of length $m$ that, with high probability, returns a $(1+\eps)$-approximation of the $F_2$ at all times of the stream. For a frequency vector of universe size $n$ and stream length $m = \poly(n)$, the algorithm uses $\tO{\frac{1}{\eps^2}}$ bits of space.
\end{theorem}
\begin{proof}
We show the robustness and space complexity guarantees of our algorithm separately. 

\textbf{Robustness.} 
We first track the adaptive interactions of the adversary with the corrector sketch in each level $i\in [H]$. Consider the transcript of symbols $\{\bot, \top\}^*$, where $\bot$ denotes that no update is made to the iterate $\bv$, while $\top$ signals that the current estimator is not accurate; this outcome is followed by an update to $\bv$ and causes the corrector sketch to release its current estimate $A_i$. Note that each adaptive query is precisely an interval query with interval $\left[(1-\eta)(P_i + Q_i), (1+\eta) (P_i + Q_i) \right]$. By \Cref{lem:bounded:updates}, we know that the outcome $\top = \{\High, \Low\}$ occurs at most $L_i \leq \O{\frac{1}{\eta^2} \log n}$ times over the course of the stream. As each corrector sketch $A_i$ is precisely the robust sketch from \Cref{lem:robust:sketch} for the adaptive query model which is accurate with probability $1-\frac{1}{\poly n}$, we know that by setting the number of rows to be $r \leq \O{\left(L_i + \frac{\sqrt{L_i}}{\eta} + \frac{1}{\eta^2}\right) \cdot \polylog\left(\frac{n}{\eta}\right)}$ and running the ADA mechanism in \Cref{alg:output}, we know that the algorithm will return a robust $(1+\eta)$-approximation to the $L_i$ adaptive queries (by \Cref{cor:conditional:release} and \Cref{rem:cond:rel:linear}). Similarly, to track the adaptive interactions with the estimator sketch, we examine the outputs produced by $P_i$ ($Q_i$ is then handled by induction). At each time, the estimator produces an estimate of $P_i$ using the sketch $\bB_i$. If $P_i + Q_i$ is accurate, then the estimator releases the \textit{same} value for $P_i$ ($\bot$), otherwise the iterate $\bv$ is updated and in the next step a \textit{re-computed} value of $P_i$ is released ($\top$). As the total number of times that $\bv$ is updated is $L_i \leq \O{\frac{1}{\eta^2} \log n}$, we again conclude that the sketch needs to be robust to only $\O{\frac{1}{\eta^2} \log n}$ adaptive queries, and thus the robustness again follows from \Cref{cor:conditional:release} and \Cref{rem:cond:rel:linear} with interval queries $\left[(1-\eta)A_i, (1+\eta) A_i \right]$. 

\textbf{Space and Bit Complexity.} We next examine the total space complexity of the algorithm. To start, note that by the above, the total number of rows in each sketch $\bB_i$ is $\O{\left(L_i + \frac{\sqrt{L_i}}{\eta} + \frac{1}{\eta^2}\right) \cdot \polylog\left(\frac{n}{\eta}\right)}$, and since $L_i \leq \O{\frac{1}{\eta^2} \log n} = \O{\frac{1}{\eps^2} \log^3 n}$, each sketch has
\[\O{\left(\frac{1}{\eps^2} \log^3 n +\frac{1}{\eps^2} \log^{2.5} n + \frac{1}{\eps^2} \log^2 n \right) \cdot \polylog\left(\frac{\log n}{\eps \delta}\right)}\]
rows. At this point, it suffices for us to bound the bit-complexity of the vectors and scalars stored throughout each active block $\calP_i$ in level $i\in [H]$ of the algorithm. To this end, consider the time frame that $\calP_i$ is active, and group the scalar multipliers ($\gamma_i$) according to which \textit{parent level} induced them. For each parent level, we have that the product of the multipliers is bounded by $M^{5/2}$ by \Cref{lem:bit:complexity}, thus the total scaling at level $i$ can be at most $M^{\O{H}}$. As we set the height of the binary recursive tree to be $H = \O{\log n}$, it follows that the total scaling is $M^{\O{\log n}}$. We remark that the values at level $i$ are also lower-bounded by $\frac{1}{\poly(n)}$, as a fresh auxiliary coordinate is added at each time-step, and this is scaled by at least $4^{-H} \cdot \delta \geq \frac{1}{\poly(n)}$ in the $i$-th level of the tree.

We now argue that the rounding error is sufficient to obtain a $(1+\eta)$-approximation at each level $i \in [H]$. Concretely, let $\by_t$ denote the vector maintained by any sketch in the hierarchy after the $t$-th time that the parent node updates its iterate. Recall that the algorithm maintains un-factored sketches of relevant vectors and rounds to $\frac{1}{\poly(n)}$-precision after each operation. Each sketched vector satisfies
\[\by_{t+1} = \gamma_t \cdot \by_{t} + \bu_t,\]
where $\bu_t$ denotes the sketched vector containing new updates after the most recent rescaling by $\gamma_t$. Let $\hat \by_{t}$ denote the rounded vector, such that 
\[\hat \by_{t+1} = \gamma_t \cdot \hat \by_{t} + \bu_t + \zeta_t .\]

Suppose the rounded version of the frequency vector incurs error $\|\zeta_t\|_{\infty} \leq h$ per operation. It then follows that 
\[\|\hat \by_t - \by_t \|_{\infty} \leq \sum_{s < t} \|\zeta_s\|_{\infty}  \left(\prod_{j = s+1}^t \gamma_j \right)\leq M^{\O{\log n}} \cdot h \cdot m,\]
where $m = \poly(n)$ denotes the stream length. Thus, by setting $h \leq \frac{1}{m \cdot M^{\O{\log n}}} \cdot \frac{1}{M^5}$, this ensures that the total rounding error is bounded by $\frac{1}{M^5}$, which is negligible compared to the norms of integer-valued vectors. Note that we choose the constant in the exponent to be sufficiently large such that this error is bounded by $\eta \cdot \delta\cdot 4^{-H}/\poly(n)$, as needed for relative accuracy at each level. Since the unscaled vectors of underlying stream updates have coordinates bounded by $\poly(n)$, we have that the total bit-complexity for each sketch is $\O{\log n \cdot \log(1/h)} = \O{\log^3 n}$.

Finally, since we store a single sketch at each level of the recursive tree of height $H = \O{\log n}$,
it follows that the algorithm uses at most $\tO{\frac{1}{\eps^2}}$ bits of space overall, as desired.

\end{proof}

\section{Adaptive Data Analysis with Bounded Universe}\label{sec:ada:bounded}
The mechanism developed in \Cref{sec:ada:gen} makes no assumption about the support of the underlying distribution on the overall population. 
In this section, we show that when the universe $U$ is finite, we can achieve better sample complexity for adaptive data analysis. 
Whereas the previous approach attempted to learn an unknown but accurate reference weight, we instead attempt to learn the population distribution $D$ itself. 

To simplify intuition (at the cost of polylogarithmic factors in the sample complexity), we use \Cref{cor:conditional:release} as a black-box corrector for an additional top level of the hierarchical estimator-corrector-learner framework. 
At this level, we maintain a set of weights $Q$ to estimate the population distribution $D$, with the prediction $Qf:=\sum_{x\in U}Q(x)f(x)$ serving as the estimator's proposed release for the true answer $Df=\sum_{x\in U}D(x)f(x)$. 
For each query $f$, we ask the corrector to check the prediction $Qf$ against a reasonably accurate estimate of $Df$.  
Following the same pattern as in \Cref{sec:ada:gen}, we release the prediction if accepted. 
Otherwise, we obtain a numerical answer from the corrector and use its correction sign to update $Q$ by multiplicative weights, which is an entropy-regularized analogue of the preceding learner, with progress measured by the KL divergence between $Q$ and the true distribution $D$,
\[\Phi(Q):=D_{\mathrm{KL}}(D\|Q)=\sum_{x\in U}D(x)\ln\frac{D(x)}{Q(x)}.\]
Relative entropy is useful here because initializing $Q$ uniformly gives $\Phi(Q)\le\log|U|$ for every population distribution $D$, even if $D$ is highly concentrated. 
The main benefit is that the number of corrections can be much smaller than the number of queries. 
Namely, we show that each correction decreases this potential by $\Omega(\alpha^2)$, and thus the learner makes at most $B=\O{\frac{\log|U|}{\alpha^2}}$ corrections. 
The bounds then follow immediately by plugging in $R+T=\O{\frac{\log|U|}{\alpha^2}}$ in the statement of \Cref{cor:conditional:release}. 

\paragraph{Parameters and mechanism.}
Fix accuracy parameter $\alpha\in(0,1)$, failure probability $\xi\in(0,1)$, and a query budget $k\ge1$. 
For $d=|U|$, we set parameters
\[\gamma=\frac{\alpha}{16},\qquad B=\min\left(k,\left\lceil\frac{4\ln d}{\alpha^2}\right\rceil\right),\]
where $\gamma$ is the required accuracy from a single instance of the mechanism in \Cref{cor:conditional:release} and $B$ represents an upper bound on the number of corrections the mechanism will ultimately need. 
Let $\calC$ be an instance \Cref{cor:conditional:release} with accuracy $\gamma$, failure probability $\xi$, and budgets $m=2k$, $R=B$, and $T=B$, i.e., it will undergo at most $2k$ interactions, and it will both correct and subsequently release a numerical value at most $B$ times. Hence, $\calC$ will halt after the $(B+1)$-th response of $\High$ or $\Low$. 

\begin{figure}
\begin{mdframed}[frametitle={Finite-universe ADA mechanism},nobreak=true]
\small
\noindent
Initialize $Q$ to the uniform distribution on $U$. 
For each query $f_t:U\to[0,1]$:
\begin{enumerate}
\item 
Compute $v_t=Qf_t$ and submit the interval request
\[\left(f_t,\left[v_t-\frac{\alpha}{2},v_t+\frac{\alpha}{2}\right]\right)\]
to $\calC$, an instance of \Cref{cor:conditional:release} with the parameters above. 
\item 
If the response is $\Accept$, output $a_t=v_t$
and proceed to the next query.
\item 
Otherwise, set $\sigma_t=+1$ for $\High$ and $\sigma_t=-1$ for $\Low$, and update
\[Q'(x)=\frac{Q(x)e^{\alpha\sigma_t f_t(x)}}{\sum_{y\in U}Q(y)e^{\alpha\sigma_t f_t(y)}},\qquad Q\leftarrow Q'.\]
\item
If $\calC$ has halted, answer the current and all remaining queries with zero, making no further requests. 
Otherwise, request a numerical answer $b_t$ for $f_t$ from $\calC$ and output $a_t=\min(1,\max(0,b_t))$.
\end{enumerate}
\end{mdframed}
\caption{Mechanism for adaptive data analysis on bounded universe.}
\label{fig:ada:bounded}
\end{figure}

\paragraph{Corrector guarantees.} 
All access to the samples occurs through the same instance $\calC$, which is never restarted. 
On every execution, the mechanism makes at most $k$ interval requests and $B$ numerical requests, hence at most $2k$ requests in total. 
Indeed, a numerical request follows a $\High$ or $\Low$ response only if $\calC$ has not halted. 
Each request to $\calC$ is determined by the previous responses, which does not have direct access to the samples (even though the responses are not independent of the samples). 
Thus, the request budgets and adaptivity assumptions of \Cref{cor:conditional:release} are satisfied.

\paragraph{Progress of the learner.}
The distribution $Q$ has full support throughout the interaction, since it is initialized to the uniform distribution, and after each subsequent multiplicative-weights update, all weights still have positive values. 
Hence, $\Phi(Q)$ remains finite.
Thus, by non-negativity of relative entropy and the fact that $Q_0$ is the uniform distribution, we have
\[\Phi(Q)\ge0,\qquad\Phi(Q_0)=\ln d+\sum_{x:D(x)>0}D(x)\ln D(x)\le\ln d.\]
We remark that as in \Cref{sec:ada:gen}, the potential is used only in the analysis, as the mechanism does not have access to either $D$ or $\Phi$. 
\begin{lemma}[Progress of a correction]
\label{lem:ada:bounded:progress}
Let $\calE$ denote the event that all responses issued by $\calC$, including any stopping response, satisfy the guarantees of \Cref{cor:conditional:release}, so that $\PPr{\calE}\ge 1-\xi$. 
Conditioned on $\calE$, every update in \Cref{fig:ada:bounded} decreases $\Phi$ by at least $\alpha^2/4$.
\end{lemma}
\begin{proof}
Let $f$ and $\sigma$ be the query and correction sign, and let $Q,Q'$ be the distributions before and after the update, respectively. 
Conditioning on the event $\calE$, the $\High$ and $\Low$ guarantees imply either $Df\ge Qf+\frac{\alpha}{2}-\gamma$ or $Df\le Qf-\frac{\alpha}{2}+\gamma$, respectively.
In either case,
\[\sigma(Df-Qf)\ge\frac{\alpha}{2}-\gamma=\frac{7\alpha}{16}.\]
Write $Z=\sum_xQ(x)e^{\alpha\sigma f(x)}$. 
Recalling that $Q'(x)=\frac{Q(x)e^{\alpha\sigma_t f_t(x)}}{\sum_{y\in U}Q(y)e^{\alpha\sigma_t f_t(y)}}=\frac{Q(x)e^{\alpha\sigma_t f_t(x)}}{Z}$ and expanding the potential, we have
\[\Phi(Q')-\Phi(Q)=-\alpha\sigma Df+\ln Z.\]
To bound the normalization term, set $Z(s)=\sum_xQ(x)e^{s\sigma f(x)}$ and define $\psi(s)=\ln Z(s)$ and $Q_s(x)=\frac{Q(x)e^{s\sigma f(x)}}{Z(s)}$. 
Differentiating the finite sums expresses the first and second
derivatives as a mean and variance under $Q_s$:
\begin{align*}
\psi'(s)&=\frac{Z'(s)}{Z(s)}=\EEx{X\sim Q_s}{\sigma f(X)},\\
\psi''(s)&=\frac{Z''(s)}{Z(s)}-\left(\frac{Z'(s)}{Z(s)}\right)^2=\operatorname{Var}_{X\sim Q_s}(\sigma f(X)).
\end{align*}
At $s=0$, we have $Z(0)=1$, so $\psi(0)=0$ and $\psi'(0)=\sigma Qf$. 
Moreover, the variance is at most the expected squared distance from the midpoint $\sigma/2$. 
Since $f(X)\in[0,1]$,
\[\psi''(s)\le\EEx{X\sim Q_s}{\left(\sigma f(X)-\frac{\sigma}{2}\right)^2}\le\frac{1}{4}.\]
Then integrating this curvature bound twice gives
\[\ln Z=\psi(\alpha)=\alpha\psi'(0)+\int_0^\alpha(\alpha-s)\psi''(s)\,ds \le\alpha\sigma Qf+\frac{\alpha^2}{8}.\]
Consequently,
\[\Phi(Q')-\Phi(Q)\le-\alpha\sigma(Df-Qf)+\frac{\alpha^2}{8}\le-\frac{5\alpha^2}{16}\le-\frac{\alpha^2}{4},\]
as desired.
\end{proof}

We now use \Cref{lem:ada:bounded:progress} to bound the number of corrections and obtain the sample complexity from \Cref{cor:conditional:release}.

\begin{restatable}[Adaptive data analysis on a bounded universe]{theorem}{thmadabounded}
\label{thm:ada:bounded}
Let $U$ be a known finite nonempty universe, and fix accuracy parameter $\alpha\in(0,1)$, failure probability $\xi\in(0,1)$, and an integer query budget $k\ge1$.
For every distribution $D$ on $U$, the mechanism in \Cref{fig:ada:bounded} answers $k$ adaptively chosen queries $f_t:U\to[0,1]$ to additive error $\alpha$, simultaneously with probability at least $1-\xi$, with sample complexity 
\[\O{\left(\frac{1}{\alpha^2} +\min\left(\frac{\sqrt{k}}{\alpha}, \frac{\sqrt{\ln|U|}}{\alpha^2}\right)\right)\log^5\left(\frac{2k}{\alpha\xi}\right)}.\]
\end{restatable}
\begin{proof}
Condition on the event $\calE$. 
We first show that $\calC$ never reaches its stopping limit. 
Since the initial potential is at most $\ln d$, \Cref{lem:ada:bounded:progress} implies that after any $c$ corrections,
\[0\le\Phi(Q)\le\ln d-\frac{c\alpha^2}{4}.\]
This bound also applies to a correction that causes $\calC$ to halt, since the corresponding update to $Q$ is performed before checking whether $\calC$ has halted.

Suppose that $\calC$ issues $B+1$ responses of $\High$ or $\Low$.
If $B=k$, this is impossible because there are at most $k$ interval requests.
Otherwise, $B=\lceil4\ln d/\alpha^2\rceil$, so setting $c=B+1$ in the preceding inequality gives
\[0\le\Phi(Q)\le\ln d-\frac{(B+1)\alpha^2}{4}<0,\]
which is a contradiction. 
Thus, $\calC$ does not halt on $\calE$, and the mechanism answers all $k$ queries without using the zero-answer fallback.

We next analyze the accuracy of the outputs of the mechanism. 
If a prediction is accepted, the interval guarantee gives
\[|a_t-Df_t|=|Qf_t-Df_t|\le\frac{\alpha}{2}+\gamma=\frac{9\alpha}{16}<\alpha.\]
Otherwise, the output $b_t$ has error at most $\gamma=\frac{\alpha}{16}$. 
Clipping $b_t$ to $[0,1]$ cannot increase its distance from $Df_t\in[0,1]$, so the released answer also has error at most $\alpha$. 
Since $\PPr{\calE}\ge1-\xi$, all answers are simultaneously accurate with probability at least $1-\xi$.

Finally, all samples are used by the single instance $\calC$.
Applying \Cref{cor:conditional:release} with the chosen parameters gives a total sample complexity of
\begin{align*}
N&=\O{\left(\frac{\sqrt{2B}}{\gamma}+\frac{1}{\gamma^2}\right)\log^5\left(\frac{2k+2B}{\gamma\xi}\right)}=\O{\left(\frac{\sqrt B}{\alpha}+\frac{1}{\alpha^2}\right)\log^5\left(\frac{2k}{\alpha\xi}\right)},
\end{align*}
where we used $\gamma=\frac{\alpha}{16}$ and $B\le k$. 
If $d=1$, then $B=0$, which gives the claimed bound.
For $d\ge2$, the assumption $\alpha<1$ gives $B=\O{\frac{\log d}{\alpha^2}}$, which then gives the desired sample complexity bounds after substituting back into $N$. 
\end{proof}

\section{AI Methodology}
We used ChatGPT Sol 6.1, Astra as part of our research: this materially affected the ideas and algorithms in \Cref{sec:ada:gen} and \Cref{sec:ada:bounded}. The improved recursive algorithm for $F_2$ was developed solely by the authors, and the authors had previously identified that an improved ADA mechanism would enable obtaining nearly-optimal bounds for $F_2$. No AI-generated text appears in the paper, as authors formulated and wrote all mathematical statements and proofs themselves. The authors take full responsibility for all content in this work. 

All results in this paper were independently proved with an end-to-end Gemini 4 harness \cite{LinWDMZM26}. See \url{https://github.com/samsonzhou/ada-stream} for the full paper and Lean verification of the main results.

\def\shortbib{0}
\bibliographystyle{alpha}
\bibliography{references}

@misc{LinWDMZM26,
      title={Stellar Colosseum: A Many-Agent Harness for Long-Horizon Research in Mathematics and Theoretical Computer Science}, 
      author={Honghao Lin and David P. Woodruff and Yuan Deng and Jieming Mao and Song Zuo and Vahab Mirrokni},
      year={2026},
      eprint={2609.15983},
      archivePrefix={arXiv},
      primaryClass={cs.AI},
      url={https://arxiv.org/abs/2609.15983} 
}

@inproceedings{HardtR10,
  author       = {Moritz Hardt and
                  Guy N. Rothblum},
  title        = {A Multiplicative Weights Mechanism for Privacy-Preserving Data Analysis},
  booktitle    = {51st Annual {IEEE} Symposium on Foundations of Computer Science, {FOCS}},
  pages        = {61--70},
  year         = {2010}
}

@inproceedings{KrishnamurthySZC03, author = {Krishnamurthy, Balachander and Sen, Subhabrata and Zhang, Yin and Chen, Yan}, title = {Sketch-based change detection: methods, evaluation, and applications}, year = {2003}, isbn = {1581137737}, publisher = {Association for Computing Machinery}, address = {New York, NY, USA}, url = {https://doi.org/10.1145/948205.948236}, doi = {10.1145/948205.948236}, booktitle = {Proceedings of the 3rd ACM SIGCOMM Conference on Internet Measurement}, pages = {234–247}, numpages = {14}, location = {Miami Beach, FL, USA}, series = {IMC '03} }

@inproceedings{AlonGMS99, author = {Alon, Noga and Gibbons, Phillip B. and Matias, Yossi and Szegedy, Mario}, title = {Tracking join and self-join sizes in limited storage}, year = {1999}, isbn = {1581130627}, publisher = {Association for Computing Machinery}, address = {New York, NY, USA}, url = {https://doi.org/10.1145/303976.303978}, doi = {10.1145/303976.303978}, booktitle = {Proceedings of the Eighteenth ACM SIGMOD-SIGACT-SIGART Symposium on Principles of Database Systems}, pages = {10–20}, numpages = {11}, location = {Philadelphia, Pennsylvania, USA}, series = {PODS '99} }

@inproceedings{Woodruff04, author = {Woodruff, David}, title = {Optimal space lower bounds for all frequency moments}, year = {2004}, isbn = {089871558X}, publisher = {Society for Industrial and Applied Mathematics}, address = {USA}, booktitle = {Proceedings of the Fifteenth Annual ACM-SIAM Symposium on Discrete Algorithms}, pages = {167–175}, numpages = {9}, location = {New Orleans, Louisiana}, series = {SODA '04} }

@INPROCEEDINGS{IndykWoodruff03,
  author={Indyk, P. and Woodruff, D.},
  booktitle={44th Annual IEEE Symposium on Foundations of Computer Science, 2003. Proceedings.}, 
  title={Tight lower bounds for the distinct elements problem}, 
  year={2003},
  volume={},
  number={},
  pages={283-288},
  doi={10.1109/SFCS.2003.1238202}}

@misc{LinSWXZ25,
      title={$L_p$ Sampling in Distributed Data Streams with Applications to Adversarial Robustness}, 
      author={Honghao Lin and Zhao Song and David P. Woodruff and Shenghao Xie and Samson Zhou},
      year={2025},
      eprint={2510.22816},
      archivePrefix={arXiv},
      primaryClass={cs.DS},
      url={https://arxiv.org/abs/2510.22816}, 
}

@misc{AssadiYazdanyar26,
      title={Fully Dynamic Algorithms for Coloring Triangle-Free Graphs}, 
      author={Sepehr Assadi and Helia Yazdanyar},
      year={2026},
      eprint={2604.20648},
      archivePrefix={arXiv},
      primaryClass={cs.DS},
      url={https://arxiv.org/abs/2604.20648}, 
}

@article{GelmanLoken14,
  title={The statistical crisis in science},
  author={Gelman, Andrew and Loken, Eric},
  journal={The best writing on mathematics (Pitici M, ed)},
  volume={102},
  pages={305--318},
  year={2016}
}

@misc{AhmadianCS25,
      title={The Cost of Compression: Tight Quadratic Black-Box Attacks on Sketches for $\ell_2$ Norm Estimation}, 
      author={Sara Ahmadian and Edith Cohen and Uri Stemmer},
      year={2025},
      eprint={2507.16345},
      archivePrefix={arXiv},
      primaryClass={cs.LG},
      url={https://arxiv.org/abs/2507.16345}, 
}

@misc{CohenStemmer26,
      title={No Escape from the Quadratic Barrier: Adaptive Attacks on Linear $\ell_2$ Sketches}, 
      author={Edith Cohen and Uri Stemmer},
      year={2026},
      eprint={2610.25008},
      archivePrefix={arXiv},
      primaryClass={cs.DS},
      url={https://arxiv.org/abs/2610.25008}, 
}

@phdthesis{Steinke16,
  author  = {Steinke, Thomas Alexander},
  title   = {Upper and Lower Bounds for Privacy and Adaptivity in Algorithmic Data Analysis},
  school  = {Harvard University},
  year    = {2016},
  type    = {Doctoral dissertation}
}

@inproceedings{LyuT25, author = {Lyu, Xin and Talwar, Kunal}, title = {Fingerprinting Codes Meet Geometry: Improved Lower Bounds for Private Query Release and Adaptive Data Analysis}, year = {2025}, isbn = {9798400715105}, publisher = {Association for Computing Machinery}, address = {New York, NY, USA}, url = {https://doi.org/10.1145/3717823.3718295}, doi = {10.1145/3717823.3718295}, booktitle = {Proceedings of the 57th Annual ACM Symposium on Theory of Computing}, pages = {2374–2385}, numpages = {12}, location = {Prague, Czechia}, series = {STOC '25} }

@misc{BassilyF16,
      title={Typical Stability}, 
      author={Raef Bassily and Yoav Freund},
      year={2016},
      eprint={1604.03336},
      archivePrefix={arXiv},
      primaryClass={cs.LG},
      url={https://arxiv.org/abs/1604.03336}, 
}

@inproceedings{RapoportCS25, author = {Rapoport, Emma and Cohen, Edith and Stemmer, Uri}, title = {Tight bounds for answering adaptively chosen concentrated queries}, year = {2026}, publisher = {Curran Associates Inc.}, address = {Red Hook, NY, USA}, url = {https://doi.org/10.52202/085713-2674}, doi = {10.52202/085713-2674}, booktitle = {Proceedings of the 39th International Conference on Neural Information Processing Systems}, articleno = {2418}, numpages = {27}, location = {San Diego, CA, United StatesMexico City, Mexico, United States}, series = {NIPS '25} }

@article{DworkR14,
  author       = {Cynthia Dwork and
                  Aaron Roth},
  title        = {The Algorithmic Foundations of Differential Privacy},
  journal      = {Found. Trends Theor. Comput. Sci.},
  volume       = {9},
  number       = {3-4},
  pages        = {211--407},
  year         = {2014}
}

@article{DesfontainesP20,
  author       = {Damien Desfontaines and
                  Bal{\'{a}}zs Pej{\'{o}}},
  title        = {SoK: Differential privacies},
  journal      = {Proc. Priv. Enhancing Technol.},
  volume       = {2020},
  number       = {2},
  pages        = {288--313},
  year         = {2020}
}

@inproceedings{VadhanW21,
  author       = {Salil P. Vadhan and
                  Tianhao Wang},
  title        = {Concurrent Composition of Differential Privacy},
  booktitle    = {Theory of Cryptography - 19th International Conference, {TCC}, Proceedings, Part {II}},
  pages        = {582--604},
  year         = {2021}
}

@inproceedings{BunS16,
  author       = {Mark Bun and
                  Thomas Steinke},
  title        = {Concentrated Differential Privacy: Simplifications, Extensions, and Lower Bounds},
  booktitle    = {Theory of Cryptography - 14th International Conference, {TCC} 2016-B, Proceedings, Part {I}},
  pages        = {635--658},
  year         = {2016}
}

@misc{JiangLY26,
title = {Turnstile Streaming Algorithms Might (Still) as Well Be Linear Sketches, for Polynomial-Length Streams},
author = {Cheng Jiang and Yinchen Liu and Huacheng Yu},
month = {April},
year = {2026},
eprint={2604.22052},
      archivePrefix={arXiv},
      primaryClass={cs.DS},
      url={https://arxiv.org/abs/2604.22052}, 
}

@misc{GribelyukLWYZ26b,
      title={Adversarial Robustness for Small Frequency Moments and a Weak Equivalence Theorem for Turnstile Streams}, 
      author={Elena Gribelyuk and Honghao Lin and David P. Woodruff and Huacheng Yu and Samson Zhou},
      year={2026},
      eprint={2607.06312},
      archivePrefix={arXiv},
      primaryClass={cs.DS},
      url={https://arxiv.org/abs/2607.06312}, 
}

@inproceedings{BravermanZ25,
  author       = {Mark Braverman and
                  Or Zamir},
  title        = {Optimality of Frequency Moment Estimation},
  booktitle    = {Proceedings of the 57th Annual {ACM} Symposium on Theory of Computing, {STOC}},
  pages        = {360--370},
  year         = {2025}
}

@inproceedings{LiNW14,
  author       = {Yi Li and
                  Huy L. Nguyen and
                  David P. Woodruff},
  title        = {Turnstile streaming algorithms might as well be linear sketches},
  booktitle    = {Symposium on Theory of Computing, {STOC}},
  pages        = {174--183},
  year         = {2014}
}

@misc{AhmadianCohen2024,
      title={Unmasking Vulnerabilities: Cardinality Sketches under Adaptive Inputs}, 
      author={Sara Ahmadian and Edith Cohen},
      year={2024},
      eprint={2405.17780},
      archivePrefix={arXiv},
      primaryClass={cs.DS},
      url={https://arxiv.org/abs/2405.17780}, 
}

@article{AlonMS99,
  author    = {Noga Alon and
               Yossi Matias and
               Mario Szegedy},
  title     = {The Space Complexity of Approximating the Frequency Moments},
  journal   = {J. Comput. Syst. Sci.},
  volume    = {58},
  number    = {1},
  pages     = {137--147},
  year      = {1999}
}

@inproceedings{HardtW13,
  author       = {Moritz Hardt and
                  David P. Woodruff},
  title        = {How robust are linear sketches to adaptive inputs?},
  booktitle    = {Symposium on Theory of Computing Conference, STOC},
  pages        = {121--130},
  year         = {2013}
}

@article{BassilyNSSSU21,
  author    = {Raef Bassily and
               Kobbi Nissim and
               Adam D. Smith and
               Thomas Steinke and
               Uri Stemmer and
               Jonathan R. Ullman},
  title     = {Algorithmic Stability for Adaptive Data Analysis},
  journal   = {{SIAM} J. Comput.},
  volume    = {50},
  number    = {3},
  year      = {2021}
}

@inproceedings{KaneNW10a,
  author       = {Daniel M. Kane and
                  Jelani Nelson and
                  David P. Woodruff},
  title        = {On the Exact Space Complexity of Sketching and Streaming Small Norms},
  booktitle    = {Proceedings of the Twenty-First Annual {ACM-SIAM} Symposium on Discrete
                  Algorithms, {SODA}},
  pages        = {1161--1178},
  year         = {2010}
}

@inproceedings{GribelyukLWYZ26,
  author       = {Elena Gribelyuk and
                  Honghao Lin and
                  David P. Woodruff and
                  Huacheng Yu and
                  Samson Zhou},
  title        = {Adversarial Robustness on Insertion-Deletion Streams},
  booktitle    = {Proceedings of the 58th Annual {ACM} Symposium on Theory of Computing, {STOC}},
  year         = {2026}
}

@inproceedings{GribelyukLWYZ25,
  author       = {Elena Gribelyuk and
                  Honghao Lin and
                  David P. Woodruff and
                  Huacheng Yu and
                  Samson Zhou},
  title        = {Lifting Linear Sketches: Optimal Bounds and Adversarial Robustness},
  booktitle    = {Proceedings of the 57th Annual {ACM} Symposium on Theory of Computing, {STOC}},
  pages        = {395--406},
  year         = {2025}
}

@inproceedings{GribelyukLWYZ24,
  author       = {Elena Gribelyuk and
                  Honghao Lin and
                  David P. Woodruff and
                  Huacheng Yu and
                  Samson Zhou},
  title        = {A Strong Separation for Adversarially Robust $L_0$ Estimation
                  for Linear Sketches},
  booktitle    = {65th {IEEE} Annual Symposium on Foundations of Computer Science, {FOCS}},
  pages        = {2318--2343},
  year         = {2024}
}

@inproceedings{ChakrabartiGS22,
  author       = {Amit Chakrabarti and
                  Prantar Ghosh and
                  Manuel Stoeckl},
  title        = {Adversarially Robust Coloring for Graph Streams},
  booktitle    = {13th Innovations in Theoretical Computer Science Conference, {ITCS}},
  pages        = {37:1--37:23},
  year         = {2022}
}

@article{Ben-EliezerJWY22,
  author       = {Omri Ben{-}Eliezer and
                  Rajesh Jayaram and
                  David P. Woodruff and
                  Eylon Yogev},
  title        = {A Framework for Adversarially Robust Streaming Algorithms},
  journal      = {J. {ACM}},
  volume       = {69},
  number       = {2},
  pages        = {17:1--17:33},
  year         = {2022}
}

@inproceedings{KaplanMNS21,
  author       = {Haim Kaplan and
                  Yishay Mansour and
                  Kobbi Nissim and
                  Uri Stemmer},
  title        = {Separating Adaptive Streaming from Oblivious Streaming Using the Bounded
                  Storage Model},
  booktitle    = {Advances in Cryptology - {CRYPTO} 2021 - 41st Annual International
                  Cryptology Conference, {CRYPTO}, Proceedings, Part {III}},
  pages        = {94--121},
  year         = {2021}
}

@inproceedings{DworkFHPRR15b,
  author       = {Cynthia Dwork and
                  Vitaly Feldman and
                  Moritz Hardt and
                  Toniann Pitassi and
                  Omer Reingold and
                  Aaron Roth},
  title        = {Generalization in Adaptive Data Analysis and Holdout Reuse},
  booktitle    = {Advances in Neural Information Processing Systems 28: Annual Conference on Neural Information Processing Systems},
  pages        = {2350--2358},
  year         = {2015}
}

@inproceedings{HassidimKMMS20,
  author       = {Avinatan Hassidim and
                  Haim Kaplan and
                  Yishay Mansour and
                  Yossi Matias and
                  Uri Stemmer},
  title        = {Adversarially Robust Streaming Algorithms via Differential Privacy},
  booktitle    = {Advances in Neural Information Processing Systems 33: Annual Conference
                  on Neural Information Processing Systems, NeurIPS},
  year         = {2020}
}

@inproceedings{Ben-EliezerEO22,
  author       = {Omri Ben{-}Eliezer and
                  Talya Eden and
                  Krzysztof Onak},
  title        = {Adversarially Robust Streaming via Dense-Sparse Trade-offs},
  booktitle    = {5th Symposium on Simplicity in Algorithms, SOSA@SODA},
  pages        = {214--227},
  year         = {2022}
}

@inproceedings{HardtU14,
  author       = {Moritz Hardt and
                  Jonathan R. Ullman},
  title        = {Preventing False Discovery in Interactive Data Analysis Is Hard},
  booktitle    = {55th {IEEE} Annual Symposium on Foundations of Computer Science, {FOCS}},
  pages        = {454--463},
  year         = {2014}
}

@inproceedings{SteinkeU15,
  author    = {Thomas Steinke and
               Jonathan R. Ullman},
  title     = {Interactive Fingerprinting Codes and the Hardness of Preventing False
               Discovery},
  booktitle = {Proceedings of The 28th Conference on Learning Theory, {COLT}},
  pages     = {1588--1628},
  year      = {2015}
}

@inproceedings{WoodruffZ21,
  author       = {David P. Woodruff and
                  Samson Zhou},
  title        = {Tight Bounds for Adversarially Robust Streams and Sliding Windows via Difference Estimators},
  booktitle    = {62nd {IEEE} Annual Symposium on Foundations of Computer Science, {FOCS}},
  pages        = {1183--1196},
  year         = {2021}
}

@inproceedings{AlokhinaB24,
  author       = {Anastasiia Alokhina and
                  Jan van den Brand},
  title        = {Fully Dynamic Shortest Path Reporting Against an Adaptive Adversary},
  booktitle    = {Proceedings of the 2024 {ACM-SIAM} Symposium on Discrete Algorithms,
                  {SODA}},
  pages        = {3027--3039},
  year         = {2024}
}

@inproceedings{BernsteinBGNSS022,
  author       = {Aaron Bernstein and
                  Jan van den Brand and
                  Maximilian Probst Gutenberg and
                  Danupon Nanongkai and
                  Thatchaphol Saranurak and
                  Aaron Sidford and
                  He Sun},
  title        = {Fully-Dynamic Graph Sparsifiers Against an Adaptive Adversary},
  booktitle    = {49th International Colloquium on Automata, Languages, and Programming,
                  {ICALP}},
  pages        = {20:1--20:20},
  year         = {2022}
}

@article{CohenNSSS24,
  author       = {Edith Cohen and
                  Jelani Nelson and
                  Tam{\'{a}}s Sarl{\'{o}}s and
                  Mihir Singhal and
                  Uri Stemmer},
  title        = {One Attack to Rule Them All: Tight Quadratic Bounds for Adaptive Queries
                  on Cardinality Sketches},
  journal      = {CoRR},
  volume       = {abs/2411.06370},
  year         = {2024}
}

@inproceedings{CohenNSS23,
  author       = {Edith Cohen and
                  Jelani Nelson and
                  Tam{\'{a}}s Sarl{\'{o}}s and
                  Uri Stemmer},
  title        = {Tricking the Hashing Trick: {A} Tight Lower Bound on the Robustness
                  of CountSketch to Adaptive Inputs},
  booktitle    = {Thirty-Seventh {AAAI} Conference on Artificial Intelligence, {AAAI}},
  pages        = {7235--7243},
  year         = {2023}
}

@inproceedings{CherapanamjeriN20,
  author       = {Yeshwanth Cherapanamjeri and
                  Jelani Nelson},
  title        = {On Adaptive Distance Estimation},
  booktitle    = {Advances in Neural Information Processing Systems 33: Annual Conference
                  on Neural Information Processing Systems, NeurIPS},
  year         = {2020}
}

@inproceedings{GoodfellowSS14,
  author       = {Ian J. Goodfellow and
                  Jonathon Shlens and
                  Christian Szegedy},
  title        = {Explaining and Harnessing Adversarial Examples},
  booktitle    = {3rd International Conference on Learning Representations, {ICLR}, Conference Track Proceedings},
  year         = {2015}
}

@inproceedings{HuangPGDA17,
  author       = {Sandy H. Huang and
                  Nicolas Papernot and
                  Ian J. Goodfellow and
                  Yan Duan and
                  Pieter Abbeel},
  title        = {Adversarial Attacks on Neural Network Policies},
  booktitle    = {5th International Conference on Learning Representations, {ICLR}, Workshop Track Proceedings},
  year         = {2017}
}

@inproceedings{AthalyeEIK18,
  author       = {Anish Athalye and
                  Logan Engstrom and
                  Andrew Ilyas and
                  Kevin Kwok},
  title        = {Synthesizing Robust Adversarial Examples},
  booktitle    = {Proceedings of the 35th International Conference on Machine Learning, {ICML}},
  pages        = {284--293},
  year         = {2018}
}

@inproceedings{DworkFHPRR15,
  author       = {Cynthia Dwork and
                  Vitaly Feldman and
                  Moritz Hardt and
                  Toniann Pitassi and
                  Omer Reingold and
                  Aaron Leon Roth},
  title        = {Preserving Statistical Validity in Adaptive Data Analysis},
  booktitle    = {Proceedings of the Forty-Seventh Annual {ACM} on Symposium on Theory
                  of Computing, {STOC}},
  pages        = {117--126},
  year         = {2015}
}

@inproceedings{DinurSWZ23,
  author       = {Itai Dinur and
                  Uri Stemmer and
                  David P. Woodruff and
                  Samson Zhou},
  title        = {On Differential Privacy and Adaptive Data Analysis with Bounded Space},
  booktitle    = {Advances in Cryptology - {EUROCRYPT} 2023 - 42nd Annual International
                  Conference on the Theory and Applications of Cryptographic Techniques, Proceedings, Part {III}},
  pages        = {35--65},
  year         = {2023}
}

@inproceedings{Wajc20,
  author    = {David Wajc},
  title     = {Rounding dynamic matchings against an adaptive adversary},
  booktitle = {Proccedings of the 52nd Annual {ACM} {SIGACT} Symposium on Theory of Computing, {STOC}},
  pages     = {194--207},
  year      = {2020}
}

@inproceedings{NissimST23,
  author       = {Kobbi Nissim and
                  Uri Stemmer and
                  Eliad Tsfadia},
  title        = {Adaptive Data Analysis in a Balanced Adversarial Model},
  booktitle    = {Advances in Neural Information Processing Systems 36: Annual Conference
                  on Neural Information Processing Systems, NeurIPS},
  year         = {2023}
}

@inproceedings{BassilyNSSSU16,
  author       = {Raef Bassily and
                  Kobbi Nissim and
                  Adam D. Smith and
                  Thomas Steinke and
                  Uri Stemmer and
                  Jonathan R. Ullman},
  title        = {Algorithmic stability for adaptive data analysis},
  booktitle    = {Proceedings of the 48th Annual {ACM} {SIGACT} Symposium on Theory
                  of Computing, {STOC}},
  pages        = {1046--1059},
  year         = {2016}
}

@inproceedings{FeldmanS18,
  author       = {Vitaly Feldman and
                  Thomas Steinke},
  title        = {Calibrating Noise to Variance in Adaptive Data Analysis},
  booktitle    = {Conference On Learning Theory, {COLT}},
  pages        = {535--544},
  year         = {2018}
}

@article{CohenGNS26,
  author       = {Edith Cohen and
                  Elena Gribelyuk and
                  Jelani Nelson and
                  Uri Stemmer},
  title        = {Adaptively Robust Resettable Streaming},
  journal      = {CoRR},
  volume       = {abs/2601.21989},
  year         = {2026}
}
\end{document}